\documentclass[11pt]{article}
\usepackage[a4paper,top=2cm,bottom=2cm,left=1.5cm,right=1.5cm]{geometry}
\usepackage[utf8]{inputenc}
\usepackage{amsmath,amsfonts,amssymb}
\usepackage{natbib,hyperref}
\usepackage{xcolor}
\usepackage{tikz}
\usepackage{array}
\usepackage{comment}
\usepackage{placeins}

\renewcommand{\d}{\mbox{d}}

\newcommand{\Esp}{\mathbb{E}}

\newcommand{\Ibb}{\mathbb{I}}
\newcommand{\Nbb}{\mathbb{N}}
\newcommand{\Pcal}{\mathcal{P}}
\newcommand{\Qcal}{\mathcal{Q}}
\renewcommand{\Pr}{\mathbb{P}}

\newcommand{\fig}{Figures}
\newcommand{\simfig}{\fig}
\newcommand{\applifig}{\fig}

\newtheorem{proposition}{Proposition}
\newcommand{\eproof}{$\blacksquare$}

\newcommand{\nodesize}{2em}
\newcommand{\edgeunit}{4*\nodesize}

\title{A Markov switching discrete-time Hawkes process: application to the monitoring of bats behavior}
\author{Anna Bonnet\footnote{Sorbonne Université, Université Paris Cité, CNRS, Laboratoire de Probabilités, Statistique et Modélisation, LPSM, F-75005 Paris, France} \footnote{\tt anna.bonnet@sorbonne-universite.fr} , St\'ephane Robin$^*$\footnote{\tt stéphane.robin@sorbonne-universite.fr}}

\begin{document}

\maketitle

\begin{abstract}
 Over the past few decades, the Hawkes process has become a popular framework for modeling temporal events thanks to its flexibility to capture different dependency structures. The objective of this work is to model call sequences emitted by bats for echolocation, whose patterns are known to change depending on the animal's activity. The novelty of the model lies in the combination of a Hawkes-type dependency from past events, as well as a latent variable that encodes changes in bat behavior.
 More precisely, we consider a discrete-time version of the Hawkes process, with an exponential kernel, where the immigration term varies according to a latent Markov chain. We prove that this model is identifiable and can be reformulated in terms of a Hidden Markov Model, with Poisson emissions. Based on these properties, we show that maximum likelihood inference of the model parameters can be performed using an EM algorithm, which involves a recursive M-step. A simulation study demonstrates the performance of our approach method for estimating the parameters, recovering the number of hidden states and classifying each bin of the trajectory. Finally, we illustrate the use of the proposed modeling to distinguish different behaviors of bats, based on the recording of their cries.

\end{abstract}

\section{Introduction\label{sec:intro}}

The Hawkes process, originally defined by \cite{Hawkes1971}, has been widely used in many fields, showing its versatility for modeling the occurrences of past-dependent events.  Among the applications, we can quote seismology \citep{Ogata1978}, genomics \citep{Reynaud_Bouret_2010}, neuroscience \citep{Reynaud2018,bonnet2023inference}, social networks \citep{Rizoiu2017}. Recently, Hawkes processes have been introduced to model animal behaviour \citep{SHD20}, which is also the application that motivates this work.

\cite{Denis2024} use a Hawkes process to model ultrasounds sequences that bats emit to echolocalize. Bats are known to emit an accumulation of calls in a short amount of time, creating a phenomenon called a `buzz', and the Hawkes process precisely describes such cluster structures. Furthermore, the number of calls that bats emit are highly influenced by their activity during the calls. In particular, bats tend to emit more frequent calls while foraging. The objective of \cite{Denis2024} is then to classify each sequence in order to recover whether the location where it was recorded is a foraging site or not.  

\begin{figure}[ht]
  \begin{center}
    \includegraphics[width=0.35\textwidth, trim=10 40 20 50, clip=]{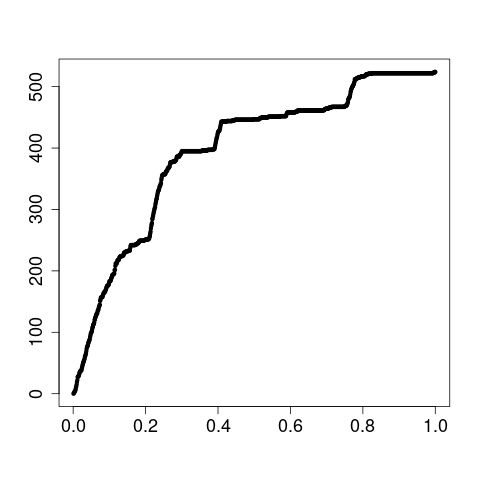} 
  \end{center}
  \caption{Example of one overnight recording of bat calls. $x$-axis = time (scaled into (0, 1)), $y$-axis = cumulated number of bat calls. \label{fig:chiro}}
\end{figure}

In the present work we study the same application with a different perspective. 
We propose an alternative modeling in which the bat's activity can change during a single sequence, therefore impacting the patterns of calls. More precisely, we propose to model the sequence of calls by a Hawkes process, the intensity of which is driven by an unobserved Markov chain that describes the changes of the bat's activity. 
Figure \ref{fig:chiro} provides an example of such a recording, on which changes in the call frequency appear clearly.
While such a switching structure is standard for Poisson processes, it remains unexplored for Hawkes processes. 

\subsection*{Our contribution}
In this paper, we define a novel modeling that we name a Markov-switching Hawkes process in discrete time, which integrates both a hidden Markov component to model the activity of the bat and a Hawkes structure to account for the dependence from the past events. 
We prove this model to be identifiable and provide a specific representation that casts it in a Hidden Markov Model framework.
This representation enables us to design a dedicated Expectation-Maximization (EM) algorithm to infer the parameters and classify time periods into latent states. 
We show on synthetic data that we both accurately estimate the parameters and recover the number of hidden states thanks to an AIC criterion. 
Finally, the procedure is illustrated on the bats' activity dataset from the Vigie-Chiro program and we demonstrate that integrating both the Hawkes structure and a Markov-switch component allows for a better fit of the data than existing approaches.  
The code is publicly available at \url{https://github.com/scj-robin/CodeHawkesDiscreteHMM}.

\subsection*{Related works}
Many works address the question of inference for Hawkes processes, using for instance least-squares \citep{Reynaud2014,Bompaire2020}, maximum likelihood \citep{Ozaki1979}, EM-algorithm \citep{Mohler2011} to name a few. In this paper, we define a discrete model that is inspired from the INAR($\infty$) process, which is also referred to as a discrete version of the Hawkes process, that is proved to converge to a continuous Hawkes process by \cite{Kirchner_SPA_16}. Estimation methods for a discretized Hawkes process include the nonparametric method from \cite{Kirchner_QF_17} or the maximum likelihood approach of \cite{Brisley2023}.

Hidden Markov Models for Poisson processes are known to be identifiable since \cite{Feller1943}. Their inference is usually addressed with an Expectation-Maximization (EM) algorithm \citep{DLR77} and have been applied in many contexts, for instance to model sismic activity \citep{Felix2022,Georgakopoulou2024} or in public health for monitoring epidemics \citep{VANKESSEL2025}.

To the best of our knowledge, the only work that introduces a Hawkes processes with a Markovian switching dynamics is \cite{Zhou2022}. However, in their case the Markovian state is observed, which was consistent with their motivating application from neuroscience. In our modeling, the activity of the bat is assumed to be unknown, as it is the case in the bat data.

Regarding the application, several works have used the proportion of buzzes in bats calls sequences to detect foraging sites \citep{GRIFFIN1960,Laforge2021}. The idea of modeling animal calls sequences as a point process to account for temporal dependence was introduced in \cite{SHD20} to describe the sound production of narwhals with autoregressive models and then developed in \cite{Denis2024} for bats monitoring. As previously mentioned, the purpose of the latter was to classify sequences but not to model and detect potential changes within a sequence.

\subsection*{Outline of the paper}
In Section \ref{sec:model}, we give a formal definition of the Markov switching Hawkes process in discrete-time and prove its identifiability. In Section \ref{sec:inference} we introduce a Hidden Markov representation of this model and derive an EM algorithm for its inference. Section \ref{sec:simuls} provides a comprehensive simulation study to assess the performances of the proposed algorithm. Eventually, we illustrate the use of the proposed modeling on bat calls sequences in Section \ref{sec:illustr}. Perspectives and future works are discussed in Section \ref{sec:discussion}.
\FloatBarrier

\section{Model\label{sec:model}}

Although the original Hawkes model \citep{Hawkes1971} is a continuous-time process, some of the literature (see for instance \cite{SEOL2015,costa2024}) focus on its discrete analogue, defined as follows : $(Y_k)_{k \geq 1}$ is a discrete-time Hawkes process if 
\begin{equation}
 Y_k \mid  (Y_h)_{1 \leq h \leq k-1} \sim \Pcal\left(\mu +  \sum_{h=1}^{k-1}\alpha_h Y_{k-h}\right)
 \label{eq:discrete_HP}
\end{equation}
Our aim is to propose an extension of this discrete version by introducing an underlying Markovian dynamics that encodes the different phases of the process and drives its parameters.

\subsection{Markov switching discrete Hawkes process}
Let us consider a discrete latent variable $(Z_k)_{k \geq 1}$ that can take $Q$ values, each of which describes an unknown state of the process. 
We assume that $(Z_k)_{k \geq 1}$ is a discrete Markov chain and we define a discrete Hawkes process conditionally on this latent variable to obtain the Markov switching discrete Hawkes process as follows:
\begin{align} \label{eq:hhmmModel}
  (Z_k)_{k \geq 1} & \sim MC(\nu, \pi), \nonumber \\
  Y_k \mid (Z_k, (Y_h)_{1 \leq h \leq k-1}) & \sim \Pcal\left(\mu_{Z_k} + \alpha \sum_{h=1}^{k-1} \beta^{h-1} Y_{k-h}\right),
\end{align}
where 
$MC(\nu, \pi)$ stands for the Markov chain over $\{1, \dots Q\}$ with initial distribution $\nu = (\nu_1, \dots \nu_Q)$ and $Q \times Q$ transition matrix $\pi$, $\mu = (\mu_1, \dots \mu_Q)$ is the vector of the baseline rates, each corresponding to one hidden state, while $\alpha$ and $\beta$ encode the dependence from the past counts $(Y_h)_{1 \leq h \leq k-1}$. The set of parameters of model \eqref{eq:hhmmModel} is hence $\theta = (\nu, \pi, \mu, \alpha,\beta)$.

Let us mention that the specific parameterization of the model comes from the choice of an exponential memory kernel that we consider here and will be detailed in the next section.
Moreover, in this modeling, one can notice that only the baseline rate depends on the hidden path while the memory coefficients $\alpha$ and $\beta$ remain constant, which means that the latent state impacts the number of events that appear spontaneously but not how the events from the past impact the present. 
One could imagine another modeling where $\alpha$ and/or $\beta$ depend on $Z_k$, this point will be further discussed in Section \ref{sec:discussion}.

\begin{figure}[ht]
  \begin{center}
    \newcommand{\bendDegree}{15}

\begin{tikzpicture}
  \node[] (Zt_3) at (-1.5*\edgeunit, 0.75*\edgeunit) {}; 
  \node[] (Zt_2) at (-1*\edgeunit,  0.75*\edgeunit) {$Z_{k-2}$}; 
  \node[] (Zt_1) at (0*\edgeunit,  0.75*\edgeunit) {$Z_{k-1}$}; 
  \node[] (Zt) at (1*\edgeunit,  0.75*\edgeunit) {$Z_{k}$}; 
  \node[] (Zt1) at (2*\edgeunit,  0.75*\edgeunit) {$Z_{k+1}$}; 
  \node[] (Zt2) at (2.5*\edgeunit,  0.75*\edgeunit) {}; 
  \node[] (Yt_3) at (-2*\edgeunit, 0) {}; 
  \node[] (Yt_2) at (-1*\edgeunit, 0) {$Y_{k-2}$}; 
  \node[] (Yt_1) at (0*\edgeunit, 0) {$Y_{k-1}$}; 
  \node[] (Yt) at (\edgeunit, 0) {$Y_{k}$}; 
  \node[] (Yt1) at (2*\edgeunit, 0) {$Y_{k+1}$}; 
  \node[] (Yt2) at (3*\edgeunit, 0) {}; 

  \draw[->,dashed] (Zt_3) -- (Zt_2); \draw[->] (Zt_2) -- (Zt_1); \draw[->] (Zt_1) -- (Zt); \draw[->] (Zt) -- (Zt1); \draw[->,dashed] (Zt1) -- (Zt2);
  \draw[->] (Zt_2) -- (Yt_2); \draw[->] (Zt_1) -- (Yt_1); \draw[->] (Zt) -- (Yt); \draw[->] (Zt1) -- (Yt1);
  \draw[->] (Zt) -- (Yt); \draw[->] (Zt1) -- (Yt1);
  
  \draw[->] (Yt_3) to[bend left=\bendDegree] (Yt_2); 
  \draw[->] (Yt_3) to[bend left=\bendDegree] (Yt_1); \draw[->] (Yt_2) to[bend left=\bendDegree] (Yt_1); 
  \draw[->] (Yt_3) to[bend right=\bendDegree] (Yt); \draw[->] (Yt_2) to[bend right=\bendDegree] (Yt);  \draw[->] (Yt_1) to[bend right=\bendDegree] (Yt); 
  \draw[->] (Yt_3) to[bend left=\bendDegree] (Yt1); \draw[->] (Yt_2) to[bend left=\bendDegree] (Yt1);  \draw[->] (Yt_1) to[bend left=\bendDegree] (Yt1); \draw[->] (Yt) to[bend left=\bendDegree] (Yt1); 
  \draw[->] (Yt_3) to[bend right=\bendDegree] (Yt2); \draw[->] (Yt_2) to[bend right=\bendDegree] (Yt2);  \draw[->] (Yt_1) to[bend right=\bendDegree] (Yt2); \draw[->] (Yt) to[bend right=\bendDegree] (Yt2); \draw[->] (Yt1) to[bend right=\bendDegree] (Yt2); 
\end{tikzpicture}
  \end{center}
  \caption{
  Directed graphical model for the process $\left((Z_k, Y_k)\right)_{k \geq 1}$ defined by model \eqref{eq:hhmmModel}. 
  \label{fig:hhmmGM-ZY}
  }.
\end{figure}
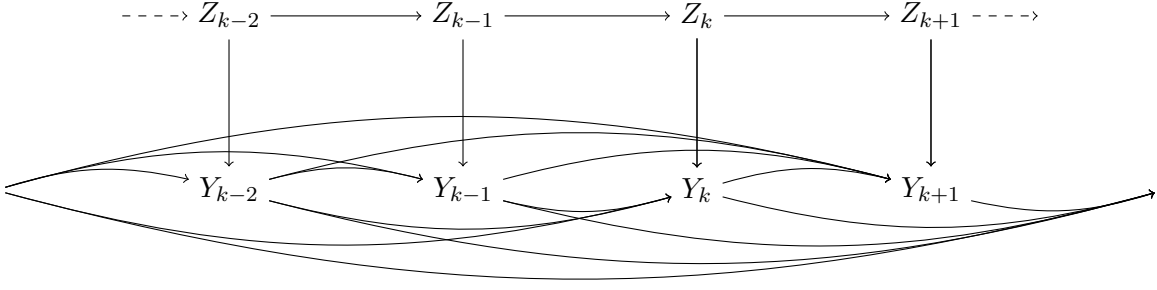

\paragraph{Graphical model.}
Figure \ref{fig:hhmmGM-ZY} displays the graphical model of the process $\left((Z_k, Y_k)\right)_{k \geq 1}$ defined by model \eqref{eq:hhmmModel}, which has an infinity memory range: the conditional distribution of $Y_k$ depends on both the current hidden state $Z_k$ and all the past observations $(Y_h)_{h < k}$.

\subsection{Connexions between continuous and discrete-time Hawkes processes} \label{sec:cont2disc}

While there are obvious similarities between the discrete and continuous versions of a Hawkes process due to a similar dependence structure from past events, the precise connection has been established by \cite{Kirchner_SPA_16}.
\cite{Kirchner_SPA_16} defines indeed a class of autoregressive point processes, called  
INAR($\infty$), which matches the discrete-time Hawkes process with infinite memory.
The main result of \cite{Kirchner_SPA_16} states that for any Hawkes process $H$, we can construct an INAR($\infty$) process such that the associated count process $H^{(\Delta)}$ on windows of size $\Delta$ converges to $H$ when $\Delta$ goes to $0$.

Let us make explicit the path from a standard Hawkes process to model \eqref{eq:hhmmModel} by deriving the expressions of the coefficients of the discrete process as functions of those of the continuous Hawkes process. Therefore, let us first consider an exponential linear Hawkes process $H$ with event times $(T_k)_{k \geq 1}$ characterized by their conditional intensity function

\begin{equation}
\lambda(t) = m + \sum_{\ell: T_\ell < t} a e^{-b(t-T_\ell)}.
\label{eq:continuous_HP}
\end{equation}

\noindent

Let us discretize the observation window $[0,T]$ into $n$ intervals $I_1, \dots, I_n$ with width $\Delta=T/n$, that is 
$$
I_k=[(k-1)\Delta ; k\Delta],
$$
for $1 \leq k \leq n$. If we denote $Y_k$ the random number of occurrences on $I_k$, then, following the decomposition in Poisson clusters \citep{Hawkes1974},
\begin{equation}
Y_k = B_k + \sum_{i=1}^{k-1} \sum_{{s} \in T_{(i)} } M_{{s}}(I_k) + R_k,
\label{eq:Y_k}
\end{equation}
where all terms of the sum are independent and:

\begin{itemize}
  \item $B_k$ is the number of occurrences coming from the Poisson baseline $m$, which means that $B_k \sim \mathcal{P}(m\Delta)$;
  \item $\sum_{i=1}^{k-1} \sum_{{s} \in T_{(i)} } M_{{s}}(I_k)$ denotes the descendants of all occurrences from past intervals $I_1, \dots, I_{k-1}$, where $T_{(i)}$ denotes the set of all occurrences that belong to $I_i=[(i-1)\Delta ;i \Delta]$. The number of descendants of each {$s \in T_{(i)}$} that belong to $I_k$ follows a Poisson distribution with parameter 
  $$
  \int_{I_k} a e^{-b(t-{s})}dt =\frac{a}{b}e^{-b(k\Delta-{s})}(e^{b\Delta}-1).
  $$
  Now if we assume that we increase the number $n$ of intervals such that their width $\Delta=T/n$ is small, then we can make the approximation {$s \simeq i\Delta$ if $s \in T_{(i)}$}, leading to
  \begin{equation}
    M_{{s}}(I_k) \simeq \Pcal\left(\frac{a (1-e^{-b\Delta})}{b} e^{-b(k-1-i)\Delta}\right) 
    \label{eq:approx_M_Ti}
  \end{equation}

  Let us notice that this approximated distribution does not depend on each ${s}$ any longer but only on the interval $I_i$ on which is located ${s}$. Then, if we introduce $\tilde{M}_{i,k}$ a sequence of independent variables distributed according to the distribution \eqref{eq:approx_M_Ti}, we obtain:
  \begin{align*}
    \sum_{i=1}^{k-1} \sum_{{s} \in T_{(i)} } M_{{s}}(I_k) & \simeq \sum_{i=1}^{k-1} \sum_{{s} \in T_{(i)} } \tilde{M}_{i,k}   =  \sum_{i=1}^{k-1} N_i \tilde{M}_{i,k} \\ 
    & \sim \Pcal\left(\sum_{i = 1}^{k-1} N_i \frac{a (1-e^{-b\Delta})}{b} e^{-b(k-1-i)\Delta} \right) = \Pcal\left(\sum_{\ell = 1}^{k-1} N_i \frac{a (1-e^{-b\Delta})}{b} e^{-b\Delta(\ell-1)} \right)
  \end{align*}
  \item $R_k$ is the number of descendants of points within $I_k$, that should be close to $0$ if the width {$\Delta$} of $I_k$ is small.
\end{itemize}

Finally, when integrating the aforementioned approximations into Equation \eqref{eq:Y_k}, we obtain an approximate distribution for $Y_k$ as follows:
\begin{equation} \label{eq:approx_Y_k}
  Y_k \mid \{Y_\ell\}_{\ell \leq k-1} \simeq \Pcal\left( \mu+ \sum_{\ell = 1}^{k} \alpha \beta^{\ell-1} Y_{k-\ell} \right),
\end{equation}
where 
\begin{align} \label{eq:cont2disc}
  \mu & = m \Delta, & \alpha & = \frac{a}b \left(1- e^{-b\Delta}\right), & \beta = e^{-b\Delta}.
\end{align} 

Then Model \eqref{eq:hhmmModel} appears as a natural extension of Model \eqref{eq:approx_Y_k} where the memory terms $\alpha$ and $\beta$ are the same while the immigration rate $\mu$ depends on a hidden path $Z$. Additionally, all parameters of Model  \eqref{eq:approx_Y_k} are explicitly derived as functions of the parameters of the standard Hawkes process \eqref{eq:continuous_HP}.

\subsection{Identifiability}

\begin{proposition} \label{prop:ident}
  The parameter set $\theta = (\nu, \pi, \mu, \alpha, \beta)$ of Model \eqref{eq:hhmmModel} is identifiable from the joint distribution of $(Y_1, Y_2, Y_3)$, that is, if for all triplet $(x, y, z)$ from $\Nbb^3$, we have that
  $
  \Pr_\theta(Y_1=x, Y_2=y, Y_3=z) = \Pr_{\theta'}(Y_1=x, Y_2=y, Y_3=z),
  $
  then $\theta = \theta'$.
\end{proposition}

\paragraph{\sl Sketch of proof.}
The detailed proof of Proposition \ref{prop:ident} is given in Appendix \ref{sec:proofIdent}. 
A main argument is that finite mixtures of Poisson distributions are identifiable. 
Based on this property, we observe that the three random variables $Y_1$, $(Y_2 \mid Y_1 = 1)$ and $(Y_3 \mid Y_1=1, Y_2 = 0)$ are each distributed according to a specific finite Poisson mixture, from which we can identify $(\nu, \mu)$, $\alpha$ and $\beta$, respectively. 
The rest of the proof consists of a deeper analysis of the joint distribution of $(Y_1, Y_2)$, from which we can identify $\pi$.
\eproof

\bigskip 
Note that the generic identifiability criterion proposed by \cite{AMR09} does not apply here because the joint distribution of $(Y_1, Y_2, Y_3)$ does not have the required product form.

\FloatBarrier

\section{Inference \label{sec:inference}}

\subsection{Hidden Markov model representation}

\paragraph{Alternative formulation of Model \eqref{eq:hhmmModel}.}
The process defined by Model \eqref{eq:hhmmModel} has indeed an infinite range memory, but we show now that a third process $(U_k)_{k \geq 1}$ can be defined so that the joint process $\left((Z_k, U_k, Y_k)\right)_{k \geq 1}$ forms a Markov chain. 

\begin{proposition} \label{prop:markov}
  Consider the discrete time process $\left((Z_k, Y_k)\right)_{k \geq 1}$ defined by Model \eqref{eq:hhmmModel}, and define $U_1 = 0$ and, for $k \geq 2$,
  \begin{align} \label{eq:hhmmUk}
    U_k & = \alpha Y_{k-1} + \beta U_{k-1}, 
  \end{align}
  then the process $\left((Z_k, U_k, Y_k)\right)_{k \geq 1}$ forms a Markov chain.
\end{proposition}

\paragraph{\sl Proof.}
  First, it is easy to check that recursion \eqref{eq:hhmmUk} gives, for $k \geq 1$, $U_k = \alpha \sum_{\ell = 1}^k \beta^{\ell-1} Y_{k-\ell}$, so the distribution of $Y_k$ in Model \ref{eq:hhmmModel} can be rewritten as
  \begin{equation} \label{eq:hhmmYk}
  Y_k \mid (Z_k, (Y_h)_{1 \leq h \leq k-1}) \sim \Pcal\left(\mu_{Z_k} + U_k\right). 
  \end{equation}

  Then, we may write the joint conditional distribution $p_\theta\left(Z_k, U_k, Y_k \mid ((Z_h, U_h, Y_h))_{1 \leq h \leq k-1}\right)$ as
  \begin{align*}
    p_\theta\left(Y_k \mid Z_k, U_k, ((Z_h, U_h, Y_h))_{1 \leq h \leq k-1}\right) 
    \; p_\theta\left(U_k \mid Z_k, ((Z_h, U_h, Y_h))_{1 \leq h \leq k-1}\right) 
    \; p_\theta\left(Z_k \mid ((Z_h, U_h, Y_h))_{1 \leq h \leq k-1}\right),
  \end{align*}
  where, 
  \begin{align*}
    p_\theta\left(Y_k \mid Z_k, U_k, ((Z_h, U_h, Y_h))_{1 \leq h \leq k-1}\right) 
    & = p_\theta\left(Y_k \mid Z_k, U_k\right) & & \text{by \eqref{eq:hhmmYk}},  \\
    p_\theta\left(U_k \mid Z_k, ((Z_h, U_h, Y_h))_{1 \leq h \leq k-1}\right) 
    & = p_\theta\left(U_k \mid U_{k-1}, Y_{k-1}\right) & & \text{by \eqref{eq:hhmmUk}}, \\
    p_\theta\left(Z_k \mid ((Z_h, U_h, Y_h))_{1 \leq h \leq k-1}\right) 
    & = p_\theta\left(Z_k \mid Z_{k-1}\right) & & \text{by \eqref{eq:hhmmModel}}, 
  \end{align*}
  which, overall, implies that
  \begin{align*}
    p_\theta\left(Z_k, U_k, Y_k \mid ((Z_h, U_h, Y_h))_{1 \leq h \leq k-1}\right)
    & = p_\theta\left(Z_k, U_k, Y_k \mid Z_{k-1}, U_{k-1}, Y_{k-1}\right).
  \end{align*}
\eproof

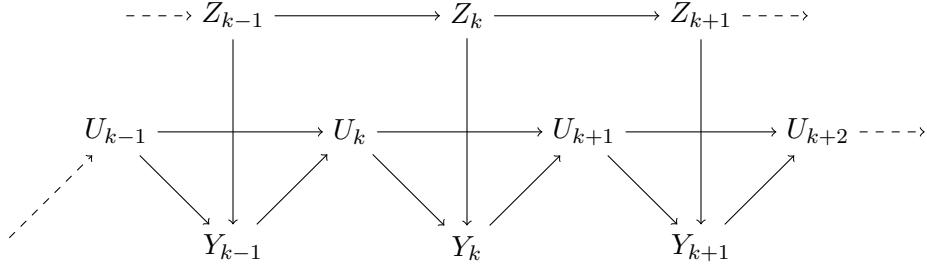
\begin{figure}[ht]
  \begin{center}
    \begin{tikzpicture}
  \node[] (Zt_2) at (-0.5*\edgeunit, \edgeunit) {}; 
  \node[] (Zt_1) at (0, \edgeunit) {$Z_{k-1}$}; 
  \node[] (Zt) at (\edgeunit, \edgeunit) {$Z_{k}$}; 
  \node[] (Zt1) at (2*\edgeunit, \edgeunit) {$Z_{k+1}$}; 
  \node[] (Zt2) at (2.5*\edgeunit, \edgeunit) {}; 
  \node[] (Ut_1) at (-0.5*\edgeunit, 0.5*\edgeunit) {$U_{k-1}$}; 
  \node[] (Ut) at (0.5*\edgeunit, 0.5*\edgeunit) {$U_{k}$}; 
  \node[] (Ut1) at (1.5*\edgeunit, 0.5*\edgeunit) {$U_{k+1}$}; 
  \node[] (Ut2) at (2.5*\edgeunit, 0.5*\edgeunit) {$U_{k+2}$}; 
  \node[] (Ut3) at (3*\edgeunit, 0.5*\edgeunit) {}; 
  \node[] (Yt_2) at (-\edgeunit, 0) {}; 
  \node[] (Yt_1) at (0, 0) {$Y_{k-1}$}; 
  \node[] (Yt) at (\edgeunit, 0) {$Y_{k}$}; 
  \node[] (Yt1) at (2*\edgeunit, 0) {$Y_{k+1}$}; 
  \node[] (Yt2) at (3*\edgeunit, 0) {}; 

  \draw[->,dashed] (Zt_2) -- (Zt_1); \draw[->] (Zt_1) -- (Zt); \draw[->] (Zt) -- (Zt1); \draw[->,dashed] (Zt1) -- (Zt2);
  \draw[->] (Zt_1) -- (Yt_1); \draw[->] (Zt) -- (Yt); \draw[->] (Zt1) -- (Yt1);
  \draw[->] (Ut_1) -- (Yt_1); \draw[->] (Ut) -- (Yt); \draw[->] (Ut1) -- (Yt1);
  \draw[->] (Ut_1) -- (Ut); \draw[->] (Ut) -- (Ut1); \draw[->] (Ut1) -- (Ut2); \draw[->,dashed] (Ut2) -- (Ut3);
  \draw[->,dashed] (Yt_2) -- (Ut_1); \draw[->] (Yt_1) -- (Ut); \draw[->] (Yt) -- (Ut1); \draw[->] (Yt1) -- (Ut2);
\end{tikzpicture}
  \end{center}
  \caption{Directed graphical model for the process $\left((Z_k, U_k, Y_k)\right)_{k \geq 1}$ defined by Equations \eqref{eq:hhmmModel}, \eqref{eq:hhmmUk} and \eqref{eq:hhmmYk}. \label{fig:hhmmGM-ZUY}}.
\end{figure}

\paragraph{Graphical model.}
The dependency structure of the process $\left((Z_k, U_k, Y_k)\right)_{k \geq 1}$ is shown in Figure \ref{fig:hhmmGM-ZUY} as its graphical model. The infinite range memory displayed in Figure \ref{fig:hhmmGM-ZY} has vanished. The $U$ coordinate of the process 'stores' all the information contained in the process's past, giving it a Markov structure.

\subsection{EM algorithm}

The Markovian nature of type process $(Z, U, Y)$ gives access to a wide range of well established procedures for the maximum likelihood inference of hidden Markov models \citep[see e.g.][]{CMR05}. The most popular one is based on the EM algorithm \citep{DLR77}, which requires the evaluation of the so-called complete likelihood and the determination of its conditional expectation given the observation $Y$.

\paragraph{Complete likelihood.}
Let us define the indicator variable $Z_{k, q} = \Ibb\{Z_k = q\}$ and denote by $\Pcal(x; \lambda) = e^{-\lambda} \lambda^x / x!$ the Poisson pdf for $x \in \Nbb$, the complete log-likelihood of the process $(Z, U, Y) = \left((Z_k, U_k, Y_k)\right)_{1 \leq k \leq n}$ then writes
\begin{align} \label{eq:hhmmLogpZUY}
  \log p_\theta\left(Z, U, Y\right)
  & = \sum_{q=1}^K Z_{1, q} \log \nu_{1q} 
  + \sum_{k=1}^n \sum_{1 \leq q, \ell \leq Q} Z_{k-1,q} Z_{k, \ell} \log \pi_{q\ell}
  + \sum_{k=1}^n \sum_{q=1}^Q Z_{k, q} \log \Pcal(Y_k; \mu_q + U_k).
\end{align}
Observe that the distribution of $U$ does not appear because $U_k$ is a deterministic function of the past, so its conditional distribution given the past is a Dirac mass at $\alpha Y_{k-1} + \beta U_{k-1}$ (or zero for $U_1$). \\
We now provide explicitly the expectation ('E') and maximization ('M') steps of the EM algorithm.

\paragraph{E step.} 
At step $h$ of the EM algorithm, the E step requires the evaluation of the conditional expectation of the complete log-likelihood \eqref{eq:hhmmLogpZUY} given the observed data $Y$ with the current estimate $\theta^{(h)} = (\nu^{(h)}, \pi^{(h)}, \mu^{(h)}, \alpha^{(h)}, \beta^{(h)})$ of the parameter set, namely
\begin{align} \label{eq:hhmmEspLogpZUY}
  \Qcal\left(\theta \mid \theta^{(h)}\right)
  & := \Esp_{\theta^{(h)}}[\log p_\theta\left(Z, U, Y\right) \mid Y] \\
  & = \sum_{q=1}^K \tau_{1, q}^{(h)} \log \nu_{1q} 
  + \sum_{k=1}^n \sum_{1 \leq q, \ell \leq Q} \eta^{(h)}_{kq\ell} \log \pi_{q\ell}
  + \sum_{k=1}^n \sum_{q=1}^Q \tau^{(h)}_{k,q}\log \Pcal(Y_k; \mu_q + U_k^{(h)}) \nonumber
\end{align}
where 
\begin{align*}
  \tau^{(h)}_{kq} & := \Esp_{\theta^{(h)}}[Z_{k,q} \mid Y] = \Pr_{\theta^{(h)}}\{Z_k = q \mid Y\}, \\
  \eta^{(h)}_{kq\ell} & := \Esp_{\theta^{(h)}}[Z_{k-1, q} Z_{k, \ell} \mid Y] = \Pr_{\theta^{(h)}}[Z_{k-1} = q, Z_k = \ell \mid Y\} & & (\text{for $k \geq 2$}).
\end{align*}
The values of $U$ at this step can be first evaluated recursively as $U^{(h)}_1 = 0$ and, for $k \geq 2$, $U^{(h)}_k = \alpha^{(h)} Y_{k-1} + \beta^{(h)} U^{(h)}_{k-1}$. 
Then, thanks Proposition \ref{prop:markov}, which ensures the hidden Markov structure of Model \eqref{eq:hhmmModel}, the conditional moments $\tau^{(h)}_{kq}$ and $\eta^{(h)}_{kq\ell}$ can be computed via the standard forward-backward recursion \citep[see, e.g.,][Section 2]{CMR05}, $\Pcal(\cdot; \mu_q + U^{(h)}_k)$ being the emission distribution at time $k$ under state $q$. 
We provide the explicit formulas in Appendix \ref{sec:forwardBackward} in Equations \eqref{eq:forward} and \eqref{eq:backward}.

\paragraph{M step.}
The M step then consists of the update of the parameter estimates as 
$$
\theta^{(h+1)}
= \arg\max_{\theta} \Qcal\left(\theta \mid \theta^{(h)}\right)
$$
where $\theta^{(h+1)}  = (\nu^{(h+1)}, \pi^{(h+1)}, \mu^{(h+1)}, \alpha^{(h+1)}, \beta^{(h+1)})$.
This step can be achieved by setting the derivatives with respect to each parameter to 0, which immediately yields, for $1 \leq q \leq Q$,
$$
\nu^{(h+1)}_q  = \tau_{1, q}^{(h)}
\qquad \text{and} \qquad   
\pi^{(h+1)}_{q\ell} = \eta^{(h)}_{kq\ell} \left/ \left( \sum_{\ell=1}^Q \eta^{(h)}_{kq\ell} \right) \right..
$$
Then, the derivatives of $\Qcal\left(\theta \mid \theta^{(h)}\right)$ with respect to $\mu$, $\alpha$ and $\beta$ can be computed recursively. Indeed, the $U_k$'s being functions of $\alpha$ and $\beta$, we have that 
\begin{align*}
  \partial_\alpha \Qcal\left(\theta \mid \theta^{(h)}\right)
  & = - \beta \; \sum_{k=1}^n \left(\partial_\alpha U_k\right) + \beta \; \sum_{k=1}^n \sum_{q} \tau_{kq} Y_k \left(\partial_\alpha U_k\right) / (\mu_q + \beta \; U_k), \\
  \partial_\beta \Qcal\left(\theta \mid \theta^{(h)}\right)
  & = - \sum_{k=1}^n (U_k + \beta \; \partial_\beta U_k) 
  + \sum_{k=1}^n \sum_{q} \tau_{kq} Y_k (U_k + \beta \; \partial_\beta U_k) / (\mu_q + \beta \; U_k), \\
  \partial_{\mu_q} \Qcal\left(\theta \mid \theta^{(h)}\right)
  & = - \sum_{k=1}^n \tau_{kq} + \sum_{k=1}^n \tau_{kq} Y_k / (\mu_q + \beta \; U_k), 
\end{align*}
with $\partial_{\alpha} U_1 = \partial_\beta U_1 = \partial_{\mu_q} U_1 = 0$ for all $q \in \{1, \dots Q\}$ and, for $k \geq 2$, using the chain rule,
\begin{align*}
  \partial_{\alpha} U_k & = Y_{k-1} + \beta \; \partial_{\alpha} U_{k-1}, & 
  \partial_{\beta} U_k & = U_{k-1} + \beta \; \partial_{\beta} U_{k-1}, & 
  \partial_{\mu_q} U_k & = 0.
\end{align*}
The maximization of $\Qcal\left(\theta \mid \theta^{(h)}\right)$ with respect to $(\alpha, \beta, \mu)$ can finally be achieved via gradient ascent.

\paragraph{Initialisation.}
As for the initial value $\theta^{(0)}$, we used maximum likelihood estimates for the homogeneous Hawkes process \citep[implemented in the {\tt hawkesbow} R package,][]{Che21} to get $\alpha^{(0)}$ and $\beta^{(0)}$, and we fitted a Poisson HMM to get $\nu^{(0)}$, $\pi^{(0)}$ and $\mu^{(0)}$.

\FloatBarrier

\section{Simulations \label{sec:simuls}}

We now present a simulation study of the accuracy of the {estimation procedure} presented in Section \ref{sec:inference}. 
More specifically, we study the influence of the intensity of the signal (number of observed events) and of the discretization step on both the accuracy of the parameter estimates and the selection of the number of hidden states.

\subsection{Simulation design}

\paragraph{Simulation model.}
Because the data to be analyzed are usually collected in continuous time (see e.g. the example from Section \ref{sec:illustr}), we simulated the continuous version of Model \ref{eq:hhmmModel} with $Q^*$ hidden states. More specifically, we simulated continuous time point processes as follows:

\begin{align} \label{eq:conthhmmModel}
  (Z(t))_{0 \leq t \leq 1} & \sim CTM(p_0, R), \\
  (H(t))_{0 \leq t \leq 1} & \sim \text{Hawkes}(\lambda), & 
  \lambda(t) & = L m_{Z(t)} + {L} \sum_{\ell: T_\ell < t} a e^{-b(t - T_\ell)}, \nonumber
\end{align}
where 
\begin{itemize}
  \item $CTM_Q(p_0, R)$ stands for the continuous-time Markov jump process over $\{1, \dots Q^*\}$ with initial distribution $p_0$ and $Q^* \times Q^*$ rate matrix $R$, 

  \item Hawkes($\lambda$) stands for the point process with conditional intensity function $\lambda$ and event times $(T_\ell)_{\ell \geq 1}$, 
  \item $m = (m_1, \dots m_{Q^*})$ stands for the vector of baseline immigration rates in states $1, \dots Q^*$ and
  \item $L > 0$ is the multiplying constant that controls the average number of events in the interval $[0, 1]$.
\end{itemize}
The set of continuous simulation parameters is therefore $(p_0, R, Lm, a, b)$, from which the corresponding set of discrete parameters $\theta = (\nu, \pi, \mu, \alpha, \beta)$ can be deduced using \eqref{eq:cont2disc} for a given time bin width $\Delta$. 

We emphasize that we are not using the most comfortable framework, since the data are not simulated according to Model \eqref{eq:hhmmModel} --~for which the proposed EM algorithm was designed~--, but under its continuous version \eqref{eq:conthhmmModel}, and then discretized. This simulation scheme makes inference more difficult, but is consistent with most practical cases.

\paragraph{Simulation parameters.}
We considered the following sets of baseline parameter values (marked with a '$^*$') for $Q^* = 1, 2, 3$ hidden states:
\begin{align*}
  Q^* & = 1: 
  & R^* & = [0], & 
  m^* & = 60; \\
  Q^* & = 2: 
  & R^* & = 25 \left[\begin{array}{rr} -1 & 1 \\ 1 & -1 \end{array}\right], & 
  m^* & = [1, 400]; \\
  Q^* & = 3:
  & R^* & = \frac{50}3 \left[\begin{array}{rrr} -2 & 1 & 1 \\ 1 & -2 & 1 \\ 1 & 1 & -2 \end{array}\right], & 
  m^* & = [1, 200, 1000];
\end{align*}
and $a^* = 40$ and $b^* = 160$ for all $Q^*$. Because of the symmetric form of the rate matrices $R$, all stationary (and initial) distributions $p_0$ are uniform over $\{1, \dots Q^*\}$. The immigration rates $m^*$ were chosen so that the mean number of events is about the same for each value of $Q^*$. The rates matrices $R^*$ were tuned so that the expected number of state changes over the interval $[0, 1]$ is 50 (when $Q^* > 1$). 

The two main tuning parameters of this simulation design are the intensity $L$ and the time bin width $\Delta$. 
For the intensity of the signal, we considered $L \in \{0.5, 1, 1.5, 2\}$. 
As for the discretization, we considered an adaptive rule, taking a number of intervals $n$proportional to the number of observed events $N = H(1)$, that is $n = CN$ sor $\Delta = 1 / n = (CN)^{-1}$, with $C \in \{0.5, 1, 2, 4\}$.

\paragraph{Data simulation.}
For each number of hidden states $Q^*$ and each intensity coefficient $L$, simulated $B = 100$ continuous-time point processes according to Model \eqref{eq:conthhmmModel}, yielding $3 \times 4 \times 100 = 1200$ simulated continuous paths. We discretized each of these paths with each discretization coefficient $C$, so to get $1200 \times 4 = 4800$ simulated discrete paths. For each  discrete path, we applied the EM algorithm described in Section \ref{sec:inference} for with $Q = 1, \dots 5$ hidden states, resulting in $4800 \times 5 = 24000$ runs of the algorithm. In each case, the EM algorithm was run until $\max_{k, q} |\tau^{(h)}_{kq} - \tau^{(h-1)}_{kq}| \leq 10^{-6}$.

\paragraph{Evaluation criteria.}
For each simulated discrete path and each number of hidden states $Q$, we recorded the estimate of the discrete parameter $\widehat{\theta} = (\widehat{\pi}, \widehat{\mu}, \widehat{\alpha}, \widehat{\beta})$ and the corresponding log-likelihood. We also we classified each time step either one by one using both the maximum a posteriori (MAP) rule: $\widehat{Z}^{MAP}_k = \arg\max_q \tau_{kq}$, or as a whole, determining the most probable hidden path $\widehat{Z}^{Vit}$ with the Viterbi algorithm: 
$$
\widehat{Z}^{Vit} = {\arg\max}_{z \in \{1 \dots Q\}^n} \; \Pr_{\widehat{\theta}}\{Z = z \mid Y\}.
$$

\begin{itemize}

  \item As for the choice of $Q$, for each discrete path $Y$ and each number of hidden states $Q$, we computed the AIC criterion \cite{Aka74}
  $$
  AIC_Q(Y) = \log p_{\widehat{\theta}_Q}(Y) - D_Q
  \qquad \text{with} \quad
  D_Q = Q^2 + 2,
  $$
  $D_Q$ being the number of independent parameters, and compared the selected $\widehat{Q} = \arg\max_Q AIC_Q(Y)$ with the true number $Q^*$. 
  \item Finally, to measure the classification accuracy, we computed the proportion of time bins for which the classification $\widehat{Z}_{kq}$ corresponds (after relabeling) to the true class $Z^*_{kq}$; $n^{-1} \sum_{k=1}^n \Ibb\{\widehat{Z}_{kq} = Z^*_{kq}\}$.
\end{itemize}

\subsection{Results}

\begin{figure}[ht]
  \begin{center}
    \includegraphics[width=\textwidth]{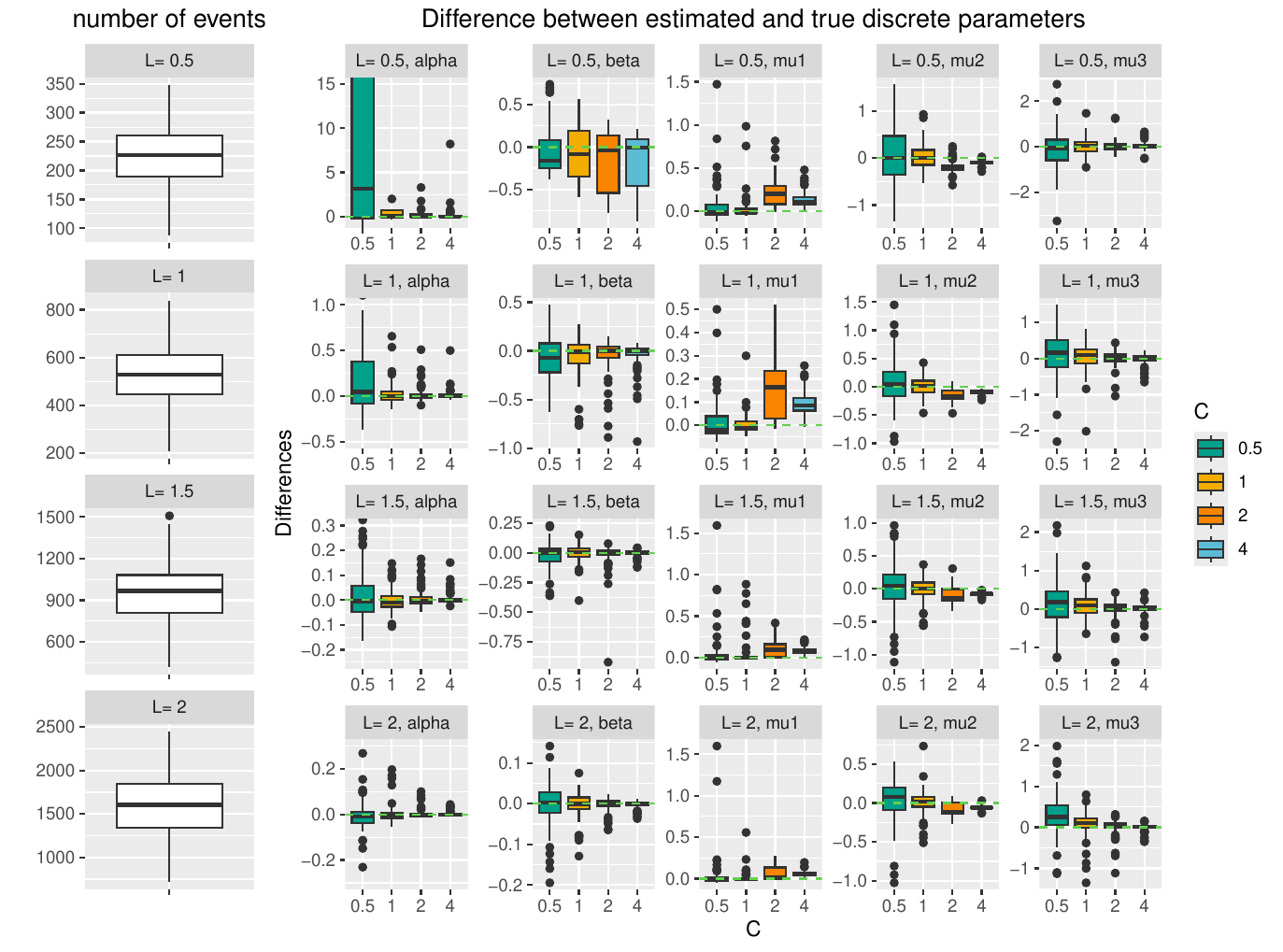}
  \end{center}
  \caption{Accuracy of the estimates of the discrete parameters $(\alpha, \beta, \mu)$ for $Q = Q^* = 3$: 
  Boxplot of the $B = 100$ differences between the parameter estimates and the true value for $\alpha$, $\beta$ and $\mu$. 
  Green dotted horizontal line at $0$.
  First column panels: number of events $n$ in the $B=100$ simulated path; 
  Second to fifth column panels: time bin width $\Delta = 1 / (CN)$, with $C = 0.5, 1, 2$ and $4$.
  First to fourth row panels: intensity coefficient $L = 0.5, 1, 1.5$ and $2$.
  \label{fig:discParmQ3}}
\end{figure}

\paragraph{Accuracy of the parameter estimates.}
Figure \ref{fig:discParmQ3} displays the results for $Q^* = 3$. We observe that accuracy of the estimates improves as the intensity parameter $L$ increases, which makes sense as this parameter controls the average number of observed events. We note also that the accuracy improves when the time bin width $\Delta$ decreases (i.e. when $C$ increases), these results being consistent with the convergence of the discrete Hawkes process toward its continuous version when the time bin width $\Delta$ goes to 0, as presented in Section \ref{sec:cont2disc}. 

The results for $Q^* = 1$ and $Q^* = 2$ are displayed in Supplementary \ref{sec:addSimuls} in Figures \ref{fig:discParmQ1} and \ref{fig:discParmQ2}, respectively, and lead to the same conclusions.

\begin{figure}[ht]
  \begin{center}
    \includegraphics[width=\textwidth]{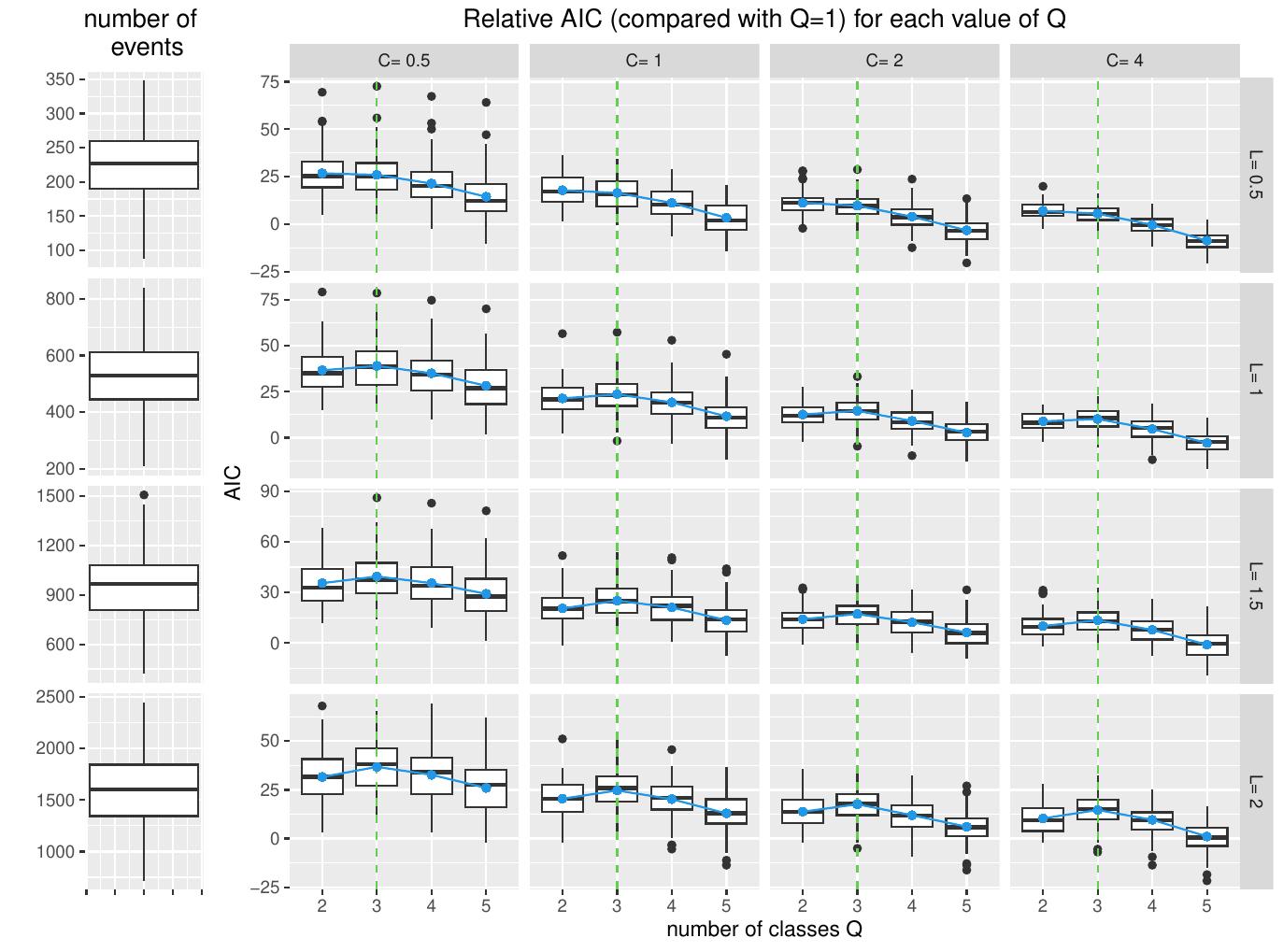}
  \end{center}
  \caption{AIC criterion for the selection of $Q$ for $Q^* = 3$:
  Boxplot of the $B = 100$ difference $AIC_Q - AIC_1$.
  Green dotted vertical line: true value $Q^*$.
  Same panels organization as in Figure \ref{fig:discParmQ3}.
  \label{fig:aicQ3}}
\end{figure}

\paragraph{Model selection.}
Figure \ref{fig:aicQ3} displays the distribution of the AIC criterion for $Q^* = 3$. Because $AIC_1$ is generally much lower than $AIC_Q$ for $Q > 1$, for the sake of clarity, we display the distribution of the difference $AIC_Q - AIC_1$. We observe a pattern similar to the one observed in terms of parameter estimation: the difference between $AIC_{Q^*}$ and $AIC_Q$ for $Q \neq Q^*$ becomes more acute as both the intensity parameter $L$ increases and the time bin width $\Delta$ decreases. 
Another visualization of the estimated value of $Q$ according to AIC criterion is given in Figure \ref{fig:Qaic_Q3} of Supplementary \ref{sec:addSimuls}, which confirms the previous comments. The results for $Q^* = 1$ and $Q^* = 2$ are also displayed in Supplementary \ref{sec:addSimuls} in Figures \ref{fig:aicQ1} and \ref{fig:aicQ2}, respectively, and yield similar conclusions. In particular, more events are required to perform model selection when $Q^*=3$ while the performance is already satisfactory with a small number of events when the number of hidden states is smaller or equal to $2$.

\begin{figure}[ht]
  \begin{center}
    \includegraphics[width=\textwidth]{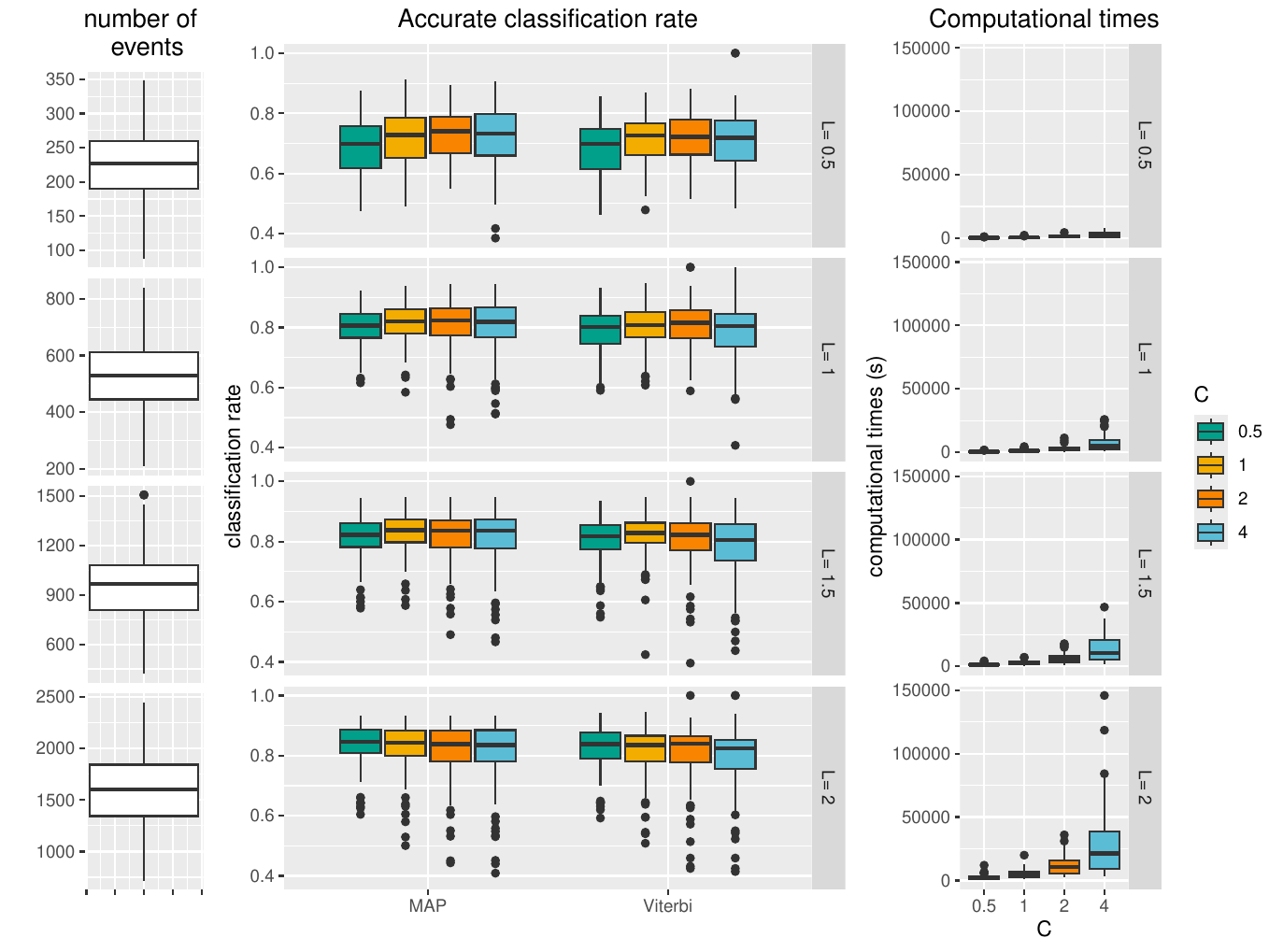}
  \end{center}
  \caption{Proportion of well classified time bins (and computational time) for $Q^* = 3$ as a function of the discretisation coefficient $C$.
  From left to right: distribution of the number of events; MAP classification rule; Viterbi classification; Total CPU time. 
  From top to bottom: $L = .5, 1, 2$ and $4$.
  \label{fig:diag-cpu-Q3}}
\end{figure}

\paragraph{Classification.}
Figure \ref{fig:diag-cpu-Q3} gives the distribution of proportion of well classified time bins with both classification strategies (MAP and Viterbi). We observe that the MAP and the Viterbi classifications yield very similar proportions, which increase when $L$ increases and also when the discretization coefficient $C$ slightly increases from $.5$ to $1$, then stabilizes for $C = 2$ and $4$. The effect of the discretization is more visible for $Q^*=2$ (see Figure \ref{fig:diag-cpu-Q2}): in such case, the more discretized the process, the better classification rate.

\paragraph{Discretization.}
All previous results suggest to use a large number $n$ of time bins but this needs to be balanced with the computational burden, as the complexity of both the E and the M step is $\Theta(n)$. 
Figure \ref{fig:diag-cpu-Q3} gives the computational times for the whole inference procedure (including the selection of $Q$). 
We observe that a high discretization ($C = 4$) yields a large increase of the computational time, while, if the gains in terms of classification and parameters estimates are obvious compared to those observed with $C=0.5$, they remain modest compared to those obtained with $C=2$.
The results for $Q^* = 2$ are given in Supplementary \ref{sec:addSimuls}, Figure \ref{fig:diag-cpu-Q2}, and lead to the same conclusions.
From our experience, taking twice as much bins as events, i.e. $C = 2$ and $\Delta = 1/ (2N)$, seems a good compromise.

\FloatBarrier

\section{Illustration \label{sec:illustr}}

We now use of the Markov-switching Hawkes process to analyze bat cries recordings, in order to capture different underlying behaviors of the animals along the night. We first demonstrate that the proposed modeling better fits the observed signal than the (homogeneous or heterogeneous) Poisson process, or the homogeneous Hawkes process. Then we focus on the number of states inferred in each cry sequence.

\begin{figure}
  \begin{center}
    \begin{tabular}{cc}
      \includegraphics[scale=0.33]{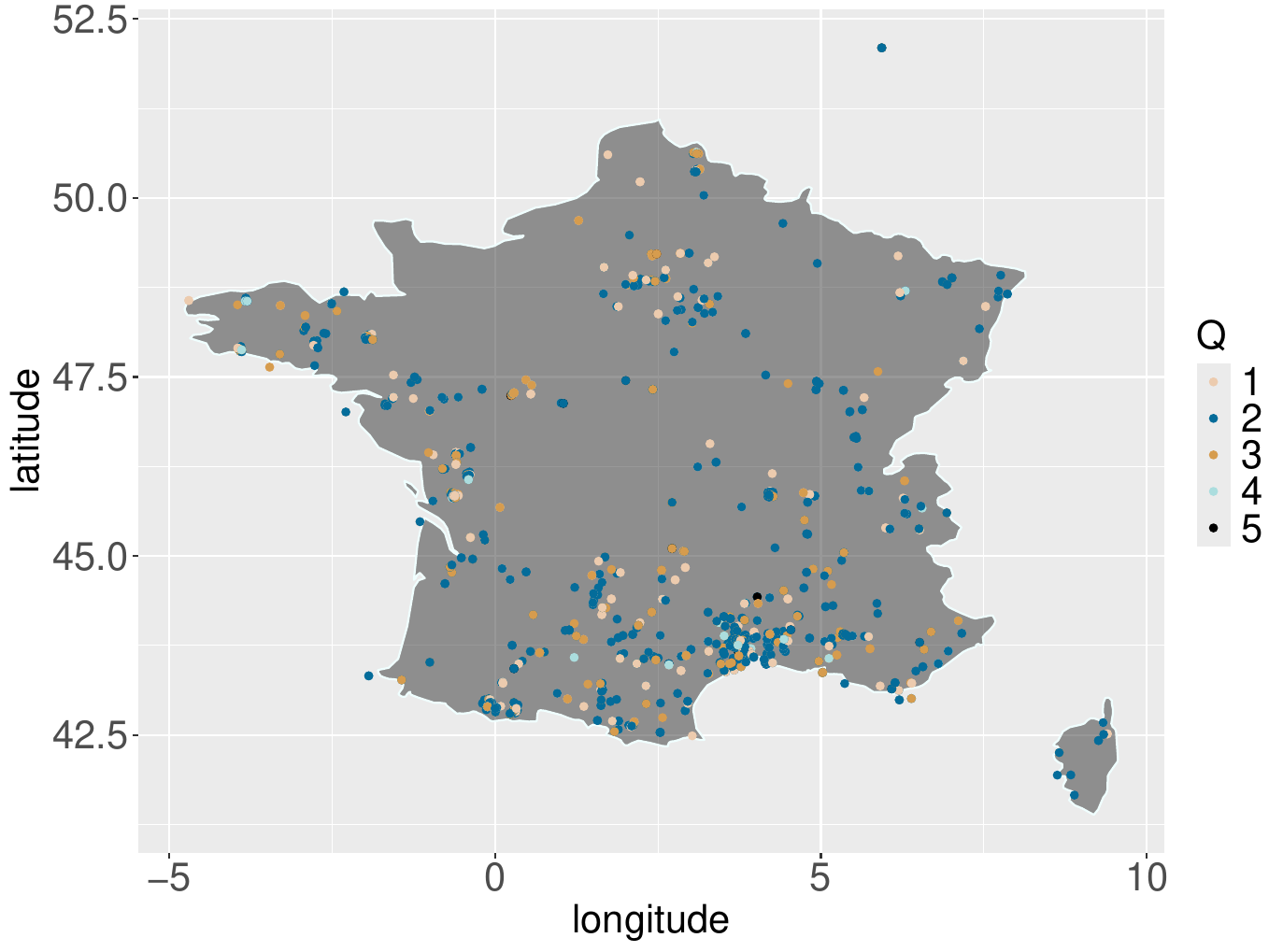} & \includegraphics[scale=0.33]{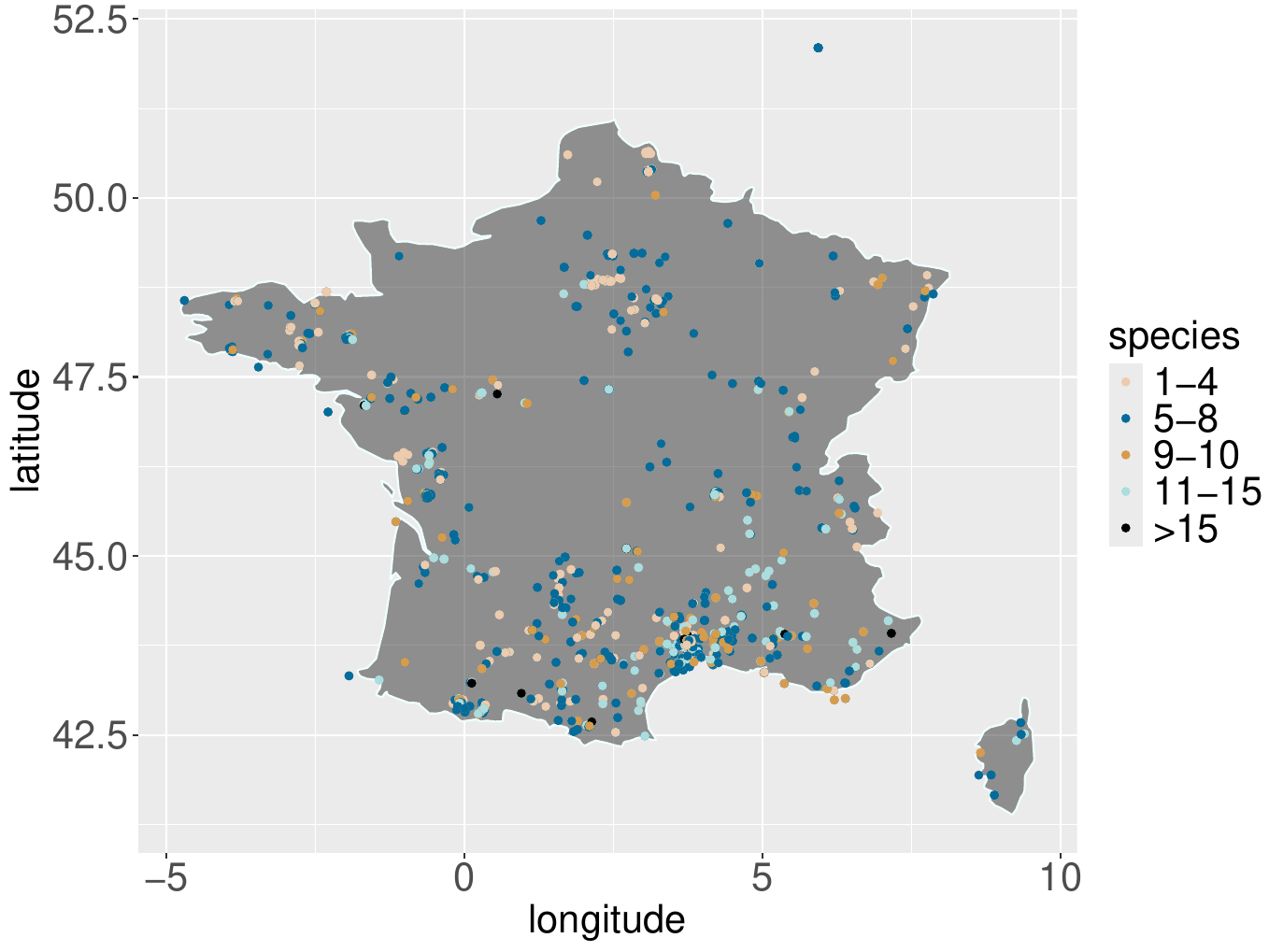} \\
      \includegraphics[scale=0.33]{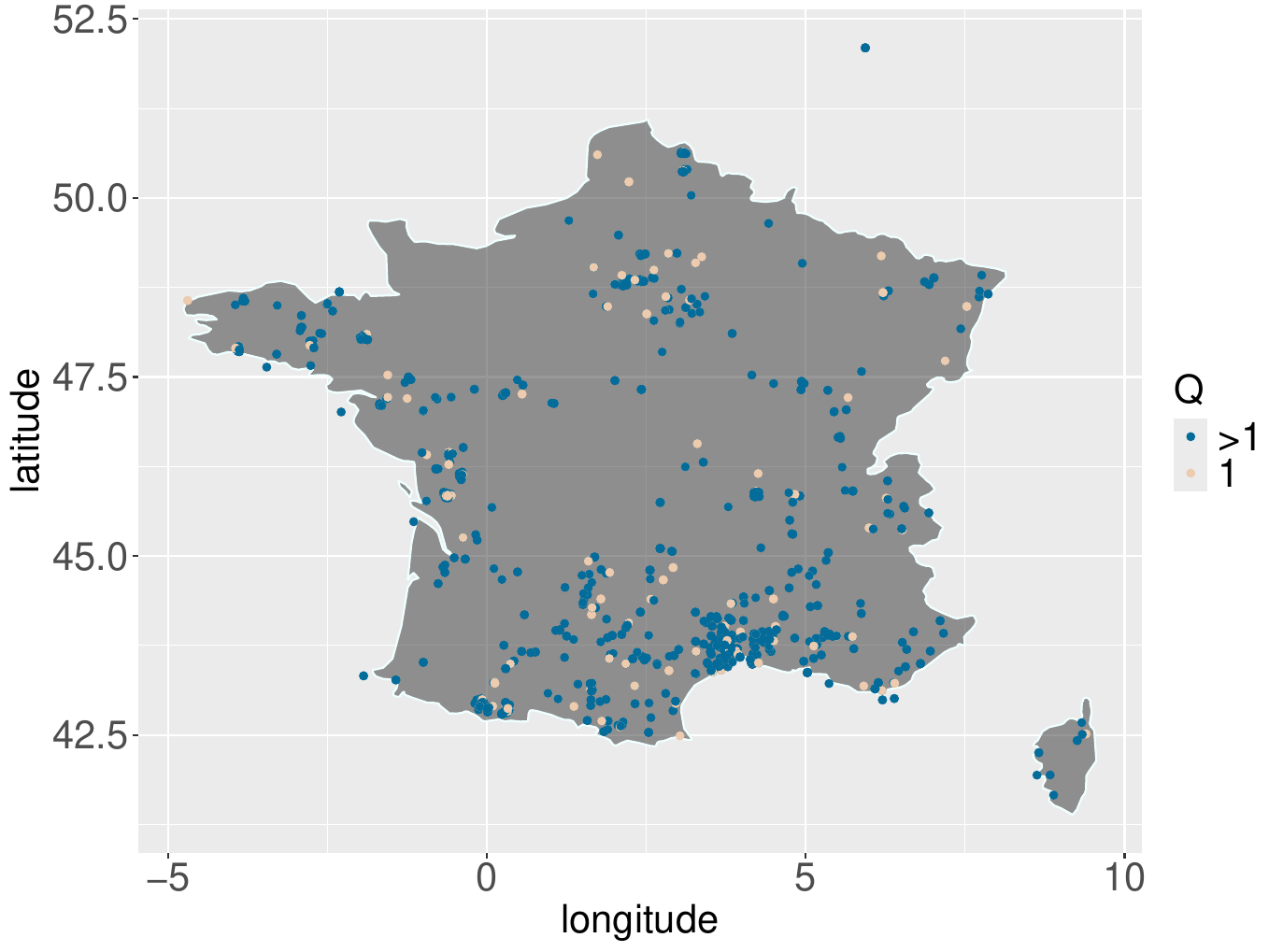} & \includegraphics[scale=0.33]{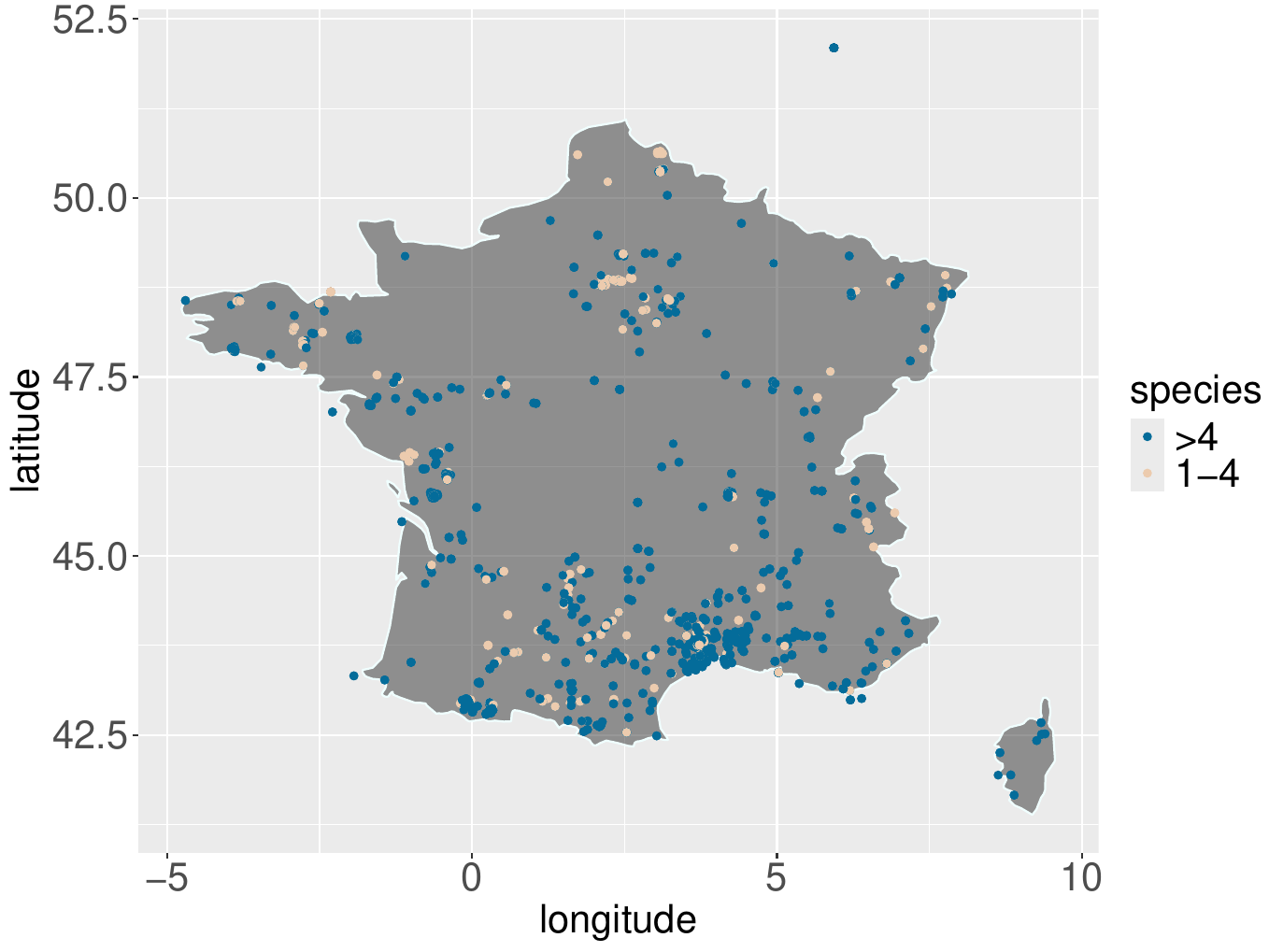} 
    \end{tabular}
  \end{center}
  \caption{Number of states $Q$ selected by our algorithm (left column) and number of species recorded (right column) for all sequences measured on each observation site. In the second row, the values have been divided into two groups: $Q=1$ vs $Q>1$ (left column) and less than $4$ species vs more than 4 species (right column).}
  \label{fig:chiro_map}
\end{figure}

\paragraph{Dataset description.}
The dataset we study comes from the Vigie-Chiro project, which is a French participatory project developed to monitor bats echolocation calls. 
In this project, Passive Acoustic Monitoring (P.A.M.) techniques are used to record ultrasonic - and therefore inaudible for human-sounds.
The dataset consists of 2354 overnight recordings of bat cries collected between October 2010 and January 2020 in 755 locations in France presented in Figure \ref{fig:chiro_map}. Each
 recording itself contains a sequence of times at which a cry occurred in the neighborhood of the device. We restricted the analysis to sequences with at least 50 cries, resulting in a set of 1555 time sequences.
A full description of the project is available at \url{https://www.vigienature.fr/fr/chauves-souris}.
\paragraph{Comparison with alternative models.} 
The motivation for developing the proposed model was to account for both a Markovian switching dynamics and a Hawkes-based dependence structure. 

In particular, the 'Hawkes-HMM' model \eqref{eq:hhmmModel} encompasses three simpler models: 
the homogeneous Poisson model (when $Q=1$ and $\alpha=0$), 
the Poisson hidden Markov model with $Q$ states ('Poisson-HMM', when $Q>1$ and $\alpha=0$), and 
the homogeneous Hawkes model (when $Q=1$).

In Table \ref{tab:best_model}, we provide, for each sequence, which of the four models is the best fit according to the AIC criterion. The conclusion is that, for most of the sequences (1144 out of 1555), our proposed model obtains the maximum AIC. Interestingly, there is significant number (353) of sequences for which the homogeneous Hawkes process remains the best fit. Finally, the two variants of the Poisson model obtain the best AIC for a small number of sequences (58). These results suggest that both the HMM part of the model and the Hawkes dependence structure are valuable for describing the data.

\begin{table}
  \begin{center}
    \begin{tabular}{>{\centering\arraybackslash}m{0.2\textwidth}|>{\centering\arraybackslash}m{0.125\textwidth}|>{\centering\arraybackslash}m{0.125\textwidth}|>{\centering\arraybackslash}m{0.125\textwidth}|m{0.15\textwidth}}
 Best model & Homogeneous Poisson & Poisson-HMM & Homogeneous Hawkes & Hawkes-HMM \\
    \hline
    Number of sequences & 34  & 24 & 353 & 1144 
    \end{tabular}
  \end{center}
  \caption{Best model according to AIC criterion for all sequences: for instance, for 34 sequences, the maximum AIC is obtained for the Poisson model. Benchmarked model: homogeneous Poisson model, Poisson with $Q$ states (``HMM-Poisson"), homogeneous Hawkes model, our proposed model (``Hawkes-HMM"). The results are obtained with the coefficient $C=2$.}
  \label{tab:best_model}
\end{table}

\begin{figure}[ht]
  \begin{center}
    \begin{tabular}{cc}
      \includegraphics[width=0.35\textwidth, trim=10 40 20 50, clip=]{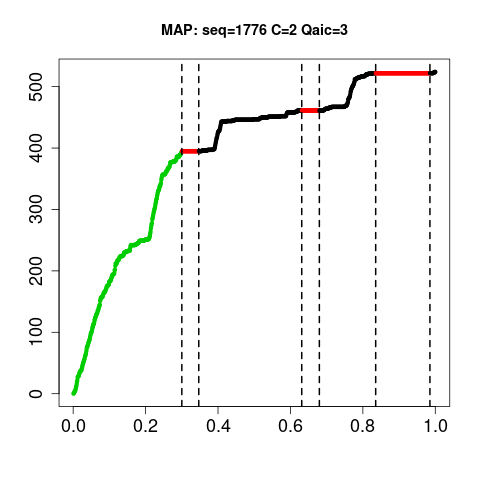} &
      \includegraphics[width=0.35\textwidth, trim=10 40 20 50, clip=]{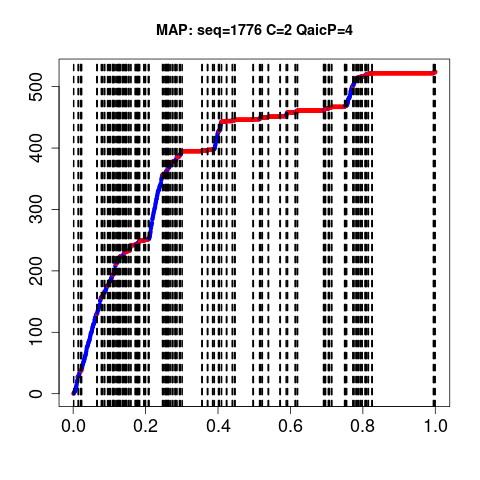} 
    \end{tabular}
  \end{center}
  \caption{Comparison between the classification provided by the Hawkes-HMM and the Poisson-HMM models: same sequence as in Figure \ref{fig:chiro}. Color = maximum a posteriori classification}
  \label{fig:chiro_hawkes_poisson_map}
\end{figure}

\paragraph{Examples of path.}
Figure \ref{fig:chiro_hawkes_poisson_map} illustrates the difference between the results  obtained with Hawkes-HMM and the Poisson HMM on the sequence presented in Figure \ref{fig:chiro}.
We clearly see that the Poisson process can not accommodate for the clustering structure of the bat calls, and compensates this lack with large number of state changes (vetical dotted lines).
We also see that the state changes detected with the Hawkes-HMM are not simply inferred when the slope varies, as slope variations also reveal clusters of calls, which are not necessarily associated with a change of behavior.
In the present case, the three inferred states could be interpreted as absence of calls (red = almost null frequency), transit (black = medium frequency) and foraging (green = high call frequency).

\paragraph{Number of hidden states.}

\begin{figure}
  \begin{center}
    \includegraphics[scale=0.35]{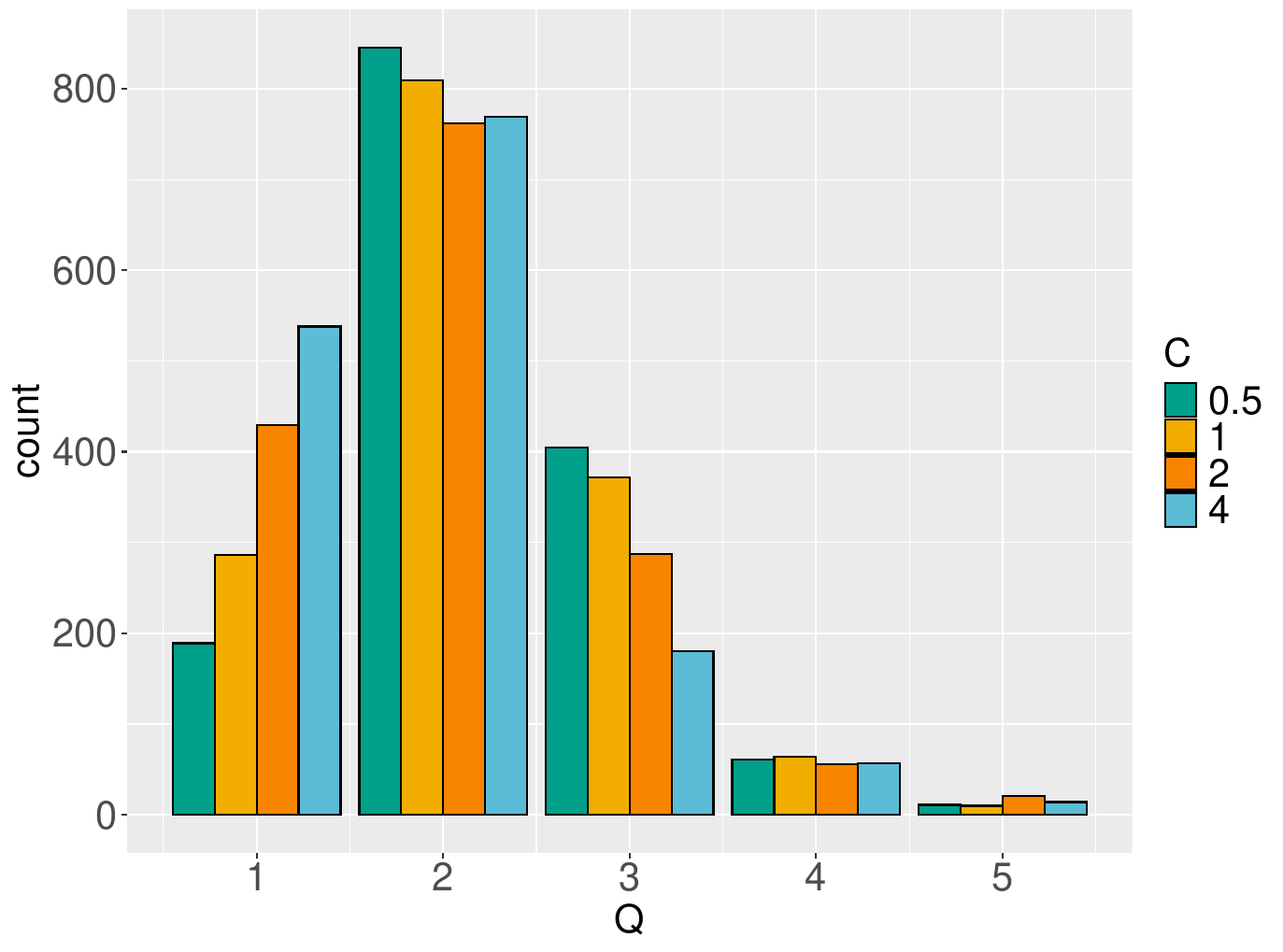}
    \includegraphics[scale=0.35]{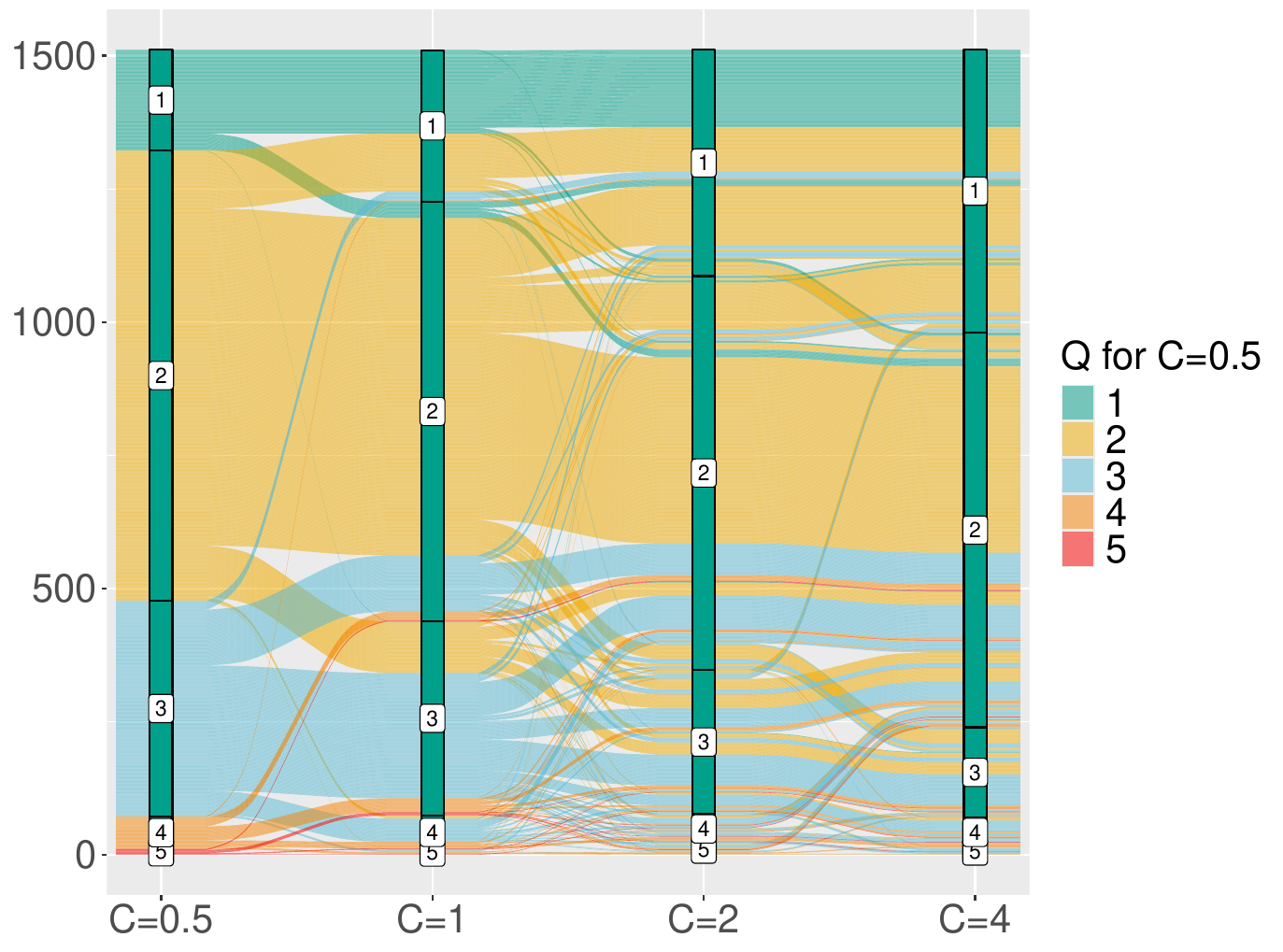}
  \end{center}
  \caption{Left: Distribution of the number of states $Q$ selected by our algorithm among all sequences, depending on the discretization parameter $C$. The sum of bars of the same color is fixed and equal to the number of sequences. \\
  Right: Alluvial plot for the number of states $Q$ selected by our algorithm depending on the discretization parameter $C$. Each flow describes the evolution of the selected value $Q$ as the discretization is thinner (i.e $C$ increases). }
  \label{fig:chiro_allu_histo}
\end{figure}

The left panel of Figure \ref{fig:chiro_allu_histo}
displays the estimated number of states $Q$ for each discretization coefficient $C$. First, we observe that for all values of $C$, the most frequent value is $Q=2$, which could correspond to the two expected behaviors, transit and foraging. Then, there is a significant number of sequences for which we estimate either $Q=1$ or $Q=3$: interestingly, the proportions are almost reversed depending on the value of $C$, which suggests that the more we discretize, the smaller number of states we detect. Finally, there is a very small number of sequences for which we estimate more than 3 states, independently on $C$. The alluvial plot presented in 
the right panel of Figure \ref{fig:chiro_allu_histo} confirms the global stability of the results according the discretization, while we also observe that the we tend to detect less changes with a thinner discretization. As suggested by the empirical results on synthetic data, we choose the intermediate value $C=2$ in the rest of the analysis.

\begin{figure}
  \begin{center}
    \includegraphics[scale=0.4]{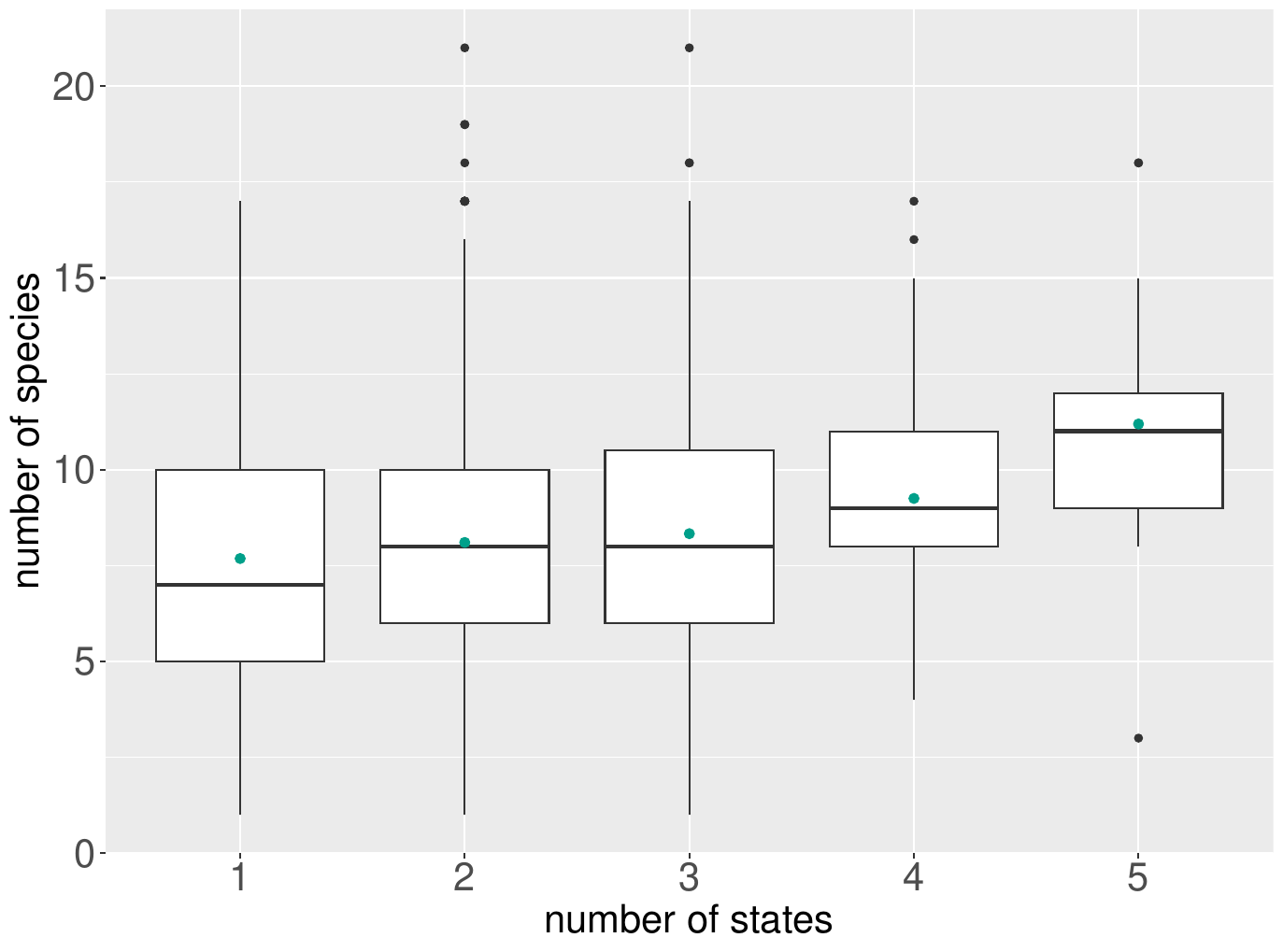}
  \end{center}
  \caption{Boxplot of the number of species for the sequences with $Q$ states selected by our algorithm. For instance, the first boxplot displays the number of species for all sequences on which the algorithm detected only one state ($Q=1$). The green point denotes the mean number of species for each value of $Q$.}
  \label{chiro_boxplot_species}
\end{figure}

\paragraph{Number of hidden states and number of species.}
Figure \ref{fig:chiro_map} shows the estimated number of hidden states $Q$ for each site, displayed on the map of France. This map is presented jointly with the map of the number of species recorded on each site. Although both maps do not overlap, there are visible similarities (in order to facilitate the comparison, we also propose a binary split of the groups). This is confirmed by Figure \ref{chiro_boxplot_species}, which shows that the value of $Q$ is positively correlated with the number of species, but the relation is far from being strict. This could suggest that different species exhibit different patterns that would be detected as different states. In order to provide a thinner interpretation of these results, it would be relevant to introduce a multivariate  process that accounts for interactions between the species. This point will be further developed in the discussion.

\FloatBarrier

\section{Discussion\label{sec:discussion}}

We have introduced a Markov-switching version of the discrete-time Hawkes process, which allows changes in the immigration rate of the process to be taken into account. In our application, these changes can be associated with modifications in animal behavior, but the framework can be adapted to a wide variety of situations in many domains, such as detecting changes in patient activity based on EEG data, or determining heterogeneous genomic regions based on pattern occurrences. We have shown that the model is identifiable and that it can be reformulated as a regular hidden Markov model, for which a genuine EM algorithm can be developed.

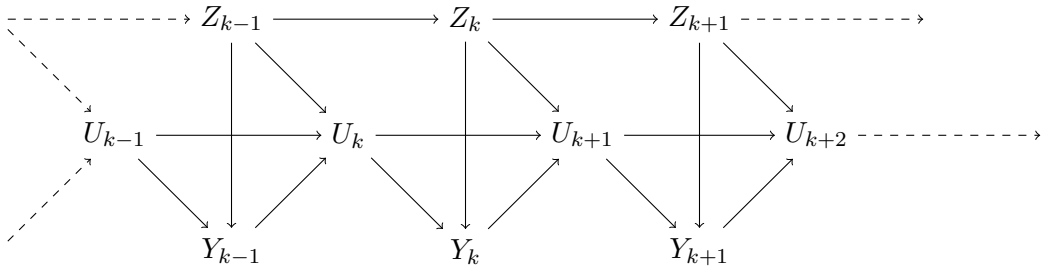
\begin{figure}[ht]
  \begin{center}
    \begin{tikzpicture}
\node[] (Zt_2) at (-\edgeunit, \edgeunit) {}; 
\node[] (Zt_1) at (0, \edgeunit) {$Z_{k-1}$}; 
\node[] (Zt) at (\edgeunit, \edgeunit) {$Z_{k}$}; 
\node[] (Zt1) at (2*\edgeunit, \edgeunit) {$Z_{k+1}$}; 
\node[] (Zt2) at (3*\edgeunit, \edgeunit) {}; 
\node[] (Ut_1) at (-0.5*\edgeunit, 0.5*\edgeunit) {$U_{k-1}$}; 
\node[] (Ut) at (0.5*\edgeunit, 0.5*\edgeunit) {$U_{k}$}; 
\node[] (Ut1) at (1.5*\edgeunit, 0.5*\edgeunit) {$U_{k+1}$}; 
\node[] (Ut2) at (2.5*\edgeunit, 0.5*\edgeunit) {$U_{k+2}$}; 
\node[] (Ut3) at (3.5*\edgeunit, 0.5*\edgeunit) {}; 
\node[] (Yt_2) at (-\edgeunit, 0) {}; 
\node[] (Yt_1) at (0, 0) {$Y_{k-1}$}; 
\node[] (Yt) at (\edgeunit, 0) {$Y_{k}$}; 
\node[] (Yt1) at (2*\edgeunit, 0) {$Y_{k+1}$}; 
\node[] (Yt2) at (3*\edgeunit, 0) {}; 

\draw[->,dashed] (Zt_2) -- (Zt_1); \draw[->] (Zt_1) -- (Zt); \draw[->] (Zt) -- (Zt1); \draw[->,dashed] (Zt1) -- (Zt2);
\draw[->] (Zt_1) -- (Yt_1); \draw[->] (Zt) -- (Yt); \draw[->] (Zt1) -- (Yt1);
\draw[->,dashed] (Zt_2) -- (Ut_1); \draw[->] (Zt_1) -- (Ut); \draw[->] (Zt) -- (Ut1); \draw[->] (Zt1) -- (Ut2);
\draw[->] (Ut_1) -- (Yt_1); \draw[->] (Ut) -- (Yt); \draw[->] (Ut1) -- (Yt1);
\draw[->] (Ut_1) -- (Ut); \draw[->] (Ut) -- (Ut1); \draw[->] (Ut1) -- (Ut2); \draw[->,dashed] (Ut2) -- (Ut3);
\draw[->,dashed] (Yt_2) -- (Ut_1); \draw[->] (Yt_1) -- (Ut); \draw[->] (Yt) -- (Ut1); \draw[->] (Yt1) -- (Ut2);
\end{tikzpicture}
  \end{center}
  \caption{
  Directed graphical model for the process $\left((Z_k, Y_k)\right)_{k \geq 1}$ , allowing the parameters $\alpha$ and $\beta$ to vary according to $Z_k$. 
  \label{fig:hhmmGM-ZY-ext}
  }
\end{figure}

\paragraph{State-dependent memory parameters.}
Several extensions of our work can be considered. First, in terms of modelling, the parameters $\alpha$ and $\beta$, which are supposed to be fixed under Model \eqref{eq:hhmmModel}, could be allowed to vary according to the hidden state $Z_k$: this would amount to replace $\alpha$ and $\beta$ with $\alpha_{Z_k}$ and $\beta_{Z_k}$, respectively. Interestingly, the model would keep its Markovian structure, as shown by its graphical model displayed in Figure \ref{fig:hhmmGM-ZY-ext}, so an EM algorithm could be designed as well. The main difficulty lies in the implementation of the update formula \eqref{eq:hhmmUk}, which then becomes $U_k = \alpha_{Z_k} Y_{k-1} + \beta_{Z_k} U_{k-1}$, and for which we would need to establish efficient recurrences to compute the conditional moment $\Qcal(\theta \mid \theta^{(h)})$ at the E step. 

\paragraph{Multivariate version.}
As mentioned in Section \ref{sec:illustr}, another natural extension of our work is to consider a multivariate version of Model \eqref{eq:hhmmModel}, to deal with events of different natures, such as cries emitted by different animals, or different species, spikes emitted by different neurons or earthquakes recorded in different locations. The multivariate Markov-switching Hawkes process would hence account for the interaction between the different processes as well as potential changes 
in terms of behavior or environment, without confounding the two.

\paragraph{Discretization and continuous time.}

As the simulations show, the choice of discretization step has an influence on the quality of inference, so the choice of $N$ remains a practical question. At this stage we only propose the heuristic rule $N = 2 n$, but this points needs to be addressed in a more theoretical way to guarantee, for example, the consistency of the deduced continuous-time parameters. Alternatively, a continuous-time version of the proposed model could be easily designed - we used it for simulations - and would benefit from a Markovian property similar to that given in Proposition \ref{prop:markov}. Here again, an EM algorithm can be designed to infer the continuous-time parameters directly, but this requires handling continuous-time integrals in an efficient way.

\subsection*{Non-exponential kernel.}
The Markovian representation given in Proposition \ref{prop:markov} does not only hold for the exponential kernel. Indeed, if the kernel function $h$ has support $(0, L\Delta)$, denoting $\alpha_\ell = \int_{(\ell-1)\Delta}^{\ell\Delta} h(s) \d s$, one may define the process $U_k = \sum_{\ell \geq 1} \alpha_\ell Y_{k-\ell}$, so that the process $\left((U_k, Y_k)\right)_{k \geq 1}$ forms a Markov chain of order $L$. A Markov switching version for this process (in which $\left((Z_k, U_k, Y_k)\right)_{k \geq 1}$ forms a Markov chain) can obviously be defined in the same way as in Model \eqref{eq:hhmmModel}, but its inference is computationally demanding as the complexity of most of the recursions involved in the EM algorithm is multiplied by $L$. Furthermore, if no assumption is made on the form of $h$, the number of parameters to estimate increases as $L$. As a consequence, the exponential model we consider seems a good balance between flexibility and computational efficiency.

\FloatBarrier

\paragraph{Acknowledgements.}
This work is part of the 2022 DAE 103 EMERGENCE(S) - PROCECO project supported by Ville de Paris. 
We are grateful to the INRAE MIGALE bioinformatics facility (MIGALE, INRAE, 2020. Migale bioinformatics Facility, doi: 10.15454/1.5572390655343293E12) for providing computing resource.
The authors are grateful to the Vigie Chiro program which provided the data (database Vigie-Nature 2025. Bat Monitoring Scheme for France. Museum national d’Histoire naturelle, Paris, France.) and in particular to
Yves Bas (Museum National d'Histoire Naturelle). The authors also thank Charlotte Dion-Blanc (Sorbonne Universit\'e, LPSM) for coordinating the ProcEco project which initiated the present work and Catherine Matias (Sorbonne Universit\'e, CNRS, LPSM) for insightful discussions.



\appendix
\newpage
\section{Appendix}


\subsection{Proof of Proposition \ref{prop:ident} \label{sec:proofIdent}}

\paragraph{Identification of $\nu$, $\mu$, $\alpha$ and $\beta$.}
The first part of the proof relies on the fact that a finite mixture of Poisson distributions $p(x) = \sum_{q=1}^Q w_q \Pcal(x; \lambda_q)$ is identifiable \citep[see][Section 3, as the set Poisson distributions is additively closed]{Tei61}.include
Based on this, we observe that the distribution of $Y_1$ is a finite Poisson mixture:
$$
\Pr_\theta\{Y_1 = x\} = \sum_{q=1}^Q \nu_q \Pcal(x; \mu_q)
$$
from which we can identify $\nu$ and $\mu$.
Then, we have that the joint distribution of $(Y_1, Y_2)$ is a mixture with $Q^2$ components:
\begin{align} \label{eq:jointY1Y2}
  \Pr_\theta\{Y_1=x, Y_2=y\} & = \sum_{1 \leq q, \ell \leq Q} \nu_q \pi_{q\ell} \phi_{q\ell}(x, y), & 
  \text{with} \quad \phi_{q\ell}(x, y) = \Pcal(x; \mu_q) \Pcal(y; \mu_\ell + \alpha x).
\end{align}
As a consequence, the distribution of $Y_2$ conditional on the event $\{Y_1 = 1\}$ is also a Poisson mixture:
$$
\Pr_\theta\{Y_2=y \mid Y_1=1\} = \sum_{\ell=1}^Q w^{(2)}_\ell \Pcal(y; \mu_\ell + \alpha),
$$
from which, as $\mu$ is known, we may identify $\alpha$ (as well as the weights $(w^{(2)}_\ell)_{1 \leq \ell \leq Q}$). Then we have that the joint distribution of $(Y_1, Y_2, Y_3)$ is a mixture with $Q^3$ components:
\begin{align*}
  \Pr_\theta\{Y_1=x, Y_2=y, Y_3=z\} & = \sum_{1 \leq q, \ell, m \leq Q} \nu_q \pi_{q\ell} \pi_{\ell m} \Pcal(x; \mu_q) \Pcal(y; \mu_\ell + \alpha x) \Pcal(y; \mu_\ell + \alpha \beta x + \alpha y),
\end{align*}
which implies that the distribution of $Y_3$ conditional on $\{Y_1=1, Y_2=0\}$ is again a Poisson mixture:
$$
\Pr_\theta\{Y_3=z \mid Y_1=1, Y_2=0\} = \sum_{m=1}^Q w^{(3)}_m \Pcal(z; \mu_m + \alpha \beta).
$$
Again, because $\mu$ and $\alpha$ are known, we may identify $\beta$ (and the weights $(w^{(3)}_m)_{1 \leq m \leq Q}$).

\paragraph{Identification of $\pi$.}
We are left with the identification of $\pi$, for which it is sufficient to prove that the mixture \eqref{eq:jointY1Y2} is identifiable. This is equivalent to show that the functions $(\phi_{q\ell})_{1 \ell q, \ell \leq Q}$ are linearly independent, that is, if there exists a set of coefficients $(c_{q\ell})_{1 \leq q, \ell \leq Q}$ such that $\sum_{1 \leq q, \ell \leq Q} c_{q\ell} \phi_{q\ell}(x, y)$ is zero for all couple $(x, y)$ from $\Nbb^2$, then all the coefficients $(c_{q\ell})_{1 \leq q, \ell \leq Q}$ are zero. \\
To this aim, we first observe that the functions $\phi_{q\ell}(x, y)$ are all proportional to $e^{-\alpha x}/(x!y!)$, which does not depend on the index $(q, \ell)$ and that the term $e^{-\mu_q - \mu_\ell}$ is a non-negative scalar, so it is equivalent to show that the functions  $(f_{q\ell})_{1 \leq q, \ell \leq Q}$, defined as $
f_{q\ell}(x, y) = \mu_q^x (\mu_\ell + \alpha x)^y$, are linearly independent. \\
Let us now suppose that there exist a set of coefficients $(d_{q\ell})_{1 \leq q, \ell \leq Q}$ such that 
$$
g(x, y) 
:= \sum_{1 \leq q, \ell \leq Q} d_{q\ell} \; f_{q\ell}(x, y)
= \sum_{1 \leq q, \ell \leq Q} d_{q\ell} \; \mu_q^x (\mu_\ell + \alpha x)^y
$$ 
is zero for all couple $(x, y)$. Now, without loss of generality, suppose that $\mu_1 < \mu_2 < \dots < \mu_Q$, and consider (as $(\mu_Q + \alpha x)^y$ is never null):
$$
\frac{g(x, y)}{(\mu_Q + \alpha x)^y}
= \sum_{q=1}^Q \sum_{\ell=1}^{Q-1} d_{q\ell} \; \mu_q^x \left(\frac{\mu_\ell + \alpha x}{\mu_Q + \alpha x}\right)^y + \sum_{q=1}^Q d_{qQ} \mu_q^x.
$$
Letting $y$ tend to infinity, the first sum vanishes, so 
$$
\lim_{y \to \infty} \frac{g(x, y)}{(\mu_Q + \alpha x)^y} = \sum_{q=1}^Q d_{qQ} \mu_q^x
$$
is zero for all $x \in \Nbb$, which implies that the $(d_{qQ})_{1 \leq q \leq Q}$ are all zero because the exponential function $\mu_q^x$ are linearly independent. \\
The rest of the proof follows by induction. Knowing that the $(d_{qQ})_{1 \leq q \leq Q}$ are all zero, we consider
$$
\frac{g(x, y)}{(\mu_{Q-1} + \alpha x)^y}
= \sum_{q=1}^Q \sum_{\ell=1}^{Q-2} d_{q\ell} \; \mu_q^x \left(\frac{\mu_\ell + \alpha x}{\mu_{Q-1} + \alpha x}\right)^y + \sum_{q=1}^Q d_{q,Q-1} \mu_q^x, 
$$
and we let $y$ tend to infinity to conclude that the $(d_{q,Q-1})_{1 \leq q \leq Q}$ are all zero as well, and so on.
\eproof

\subsection{Forward-backward recursion formulas \label{sec:forwardBackward}}

We derive here the recursion formulas for the E step of the EM algorithm presented in Section \ref{sec:inference}. For the sake of clarity, we drop the iteration superscript $(h)$. The only specificity with respect to the standard forward-backward recursion for hidden Markov models \citep[see][Section 2]{CMR05} lies in the justification of the forward recursion.

\paragraph{Forward recursion.} We define $Y_1^k = (Y_1, \dots Y_k)$ the set of observations up to time $k$ and $F_{kq}$ the conditional probability to be in state $q$ given $Y_1^k$: $F_{kq} := \Pr_\theta\{Z_k = q \mid Y_1^k\}$. 
Because $U_1 = 0$, the recursion is initialized with $F_{1q}  = \nu_q \Pcal(Y_1; \mu_q) / p_\theta(Y_1)$, where $p_\theta(Y_1) = \sum_{\ell=1}^Q \nu_\ell \Pcal(Y_1; \mu_\ell)$.
Then, for $k \geq 2$, we may decompose $F_{kq}$ as
\begin{align*}
  \Pr_\theta\{Z_k=\ell \mid Y_1^k\} 
  & = \sum_{q=1}^Q \Pr_\theta\{Z_{k-1}=q, Z_k=\ell \mid Y_1^k\} 
  = \frac1{p_\theta(Y_1^k)} \sum_{q=1}^Q \Pr_\theta\{Z_{k-1}=q, Z_k=\ell, Y_1^{k-1}, Y_k\} \\
  & = \frac1{p_\theta(Y_1^k)} \sum_{q=1}^Q \Pr_\theta\{Z_{k-1}=q, Z_k=\ell, U_k, Y_1^{k-1}, Y_k\} 
\end{align*}
because $U_k$ is a deterministic function of the past $Y_1^k$. Now, using iterative conditioning, we have that
\begin{align*}
  & \Pr_\theta\{Z_{k-1}=q, Z_k=\ell, U_k, Y_1^{k-1}, Y_k\} \\
  & = p_\theta(Y_1^{k-1})
  \Pr_\theta\{Z_{k-1}=q \mid Y_1^{k-1}\} 
  \Pr_\theta\{Z_k=\ell \mid Z_{k-1}=q, Y_1^{k-1}\} 
  \Pr_\theta\{Y_k \mid Z_{k-1}=q, Z_k=\ell, U_k, Y_1^{k-1}\} 
\end{align*}
where 
\begin{itemize}
  \item $\Pr_\theta\{Z_{k-1}=q \mid Y_1^{k-1}\} = F_{k-1, q}$ by definition of it, 
  \item $\Pr_\theta\{Z_k=\ell \mid Z_{k-1}=q, Y_1^{k-1}\} = \Pr_\theta\{Z_k=\ell \mid Z_{k-1}=q\} = \pi_{q\ell}$ by definition of Model \ref{eq:hhmmModel} and 
  \item $\Pr_\theta\{Y_k \mid Z_{k-1}=q, Z_k=\ell, U_k, Y_1^{k-1}\} = \Pr_\theta\{Y_k \mid Z_k=\ell, U_k\} = \Pcal(Y_k; \mu_q + U_k)$ thanks to Proposition \ref{prop:markov}.
\end{itemize}
Overall, we get for $k \geq 2$:
\begin{equation} \label{eq:forward}
F_{k\ell}
= 
\frac{p_\theta(Y_1^{k-1})}{p_\theta(Y_1^k)} \sum_{q=1}^Q F_{k-1, q} \pi_{q\ell} \Pcal(Y_k; \mu_q + U_k)
\propto 
\sum_{q=1}^Q F_{k-1, q} \pi_{q\ell} \Pcal(Y_k; \mu_q + U_k).
\end{equation}
The specificity of this recursion is that the parameter of the emission distribution does depend on the time $k$, through the variable $U_k$. \\
We remind that the normalizing ratio $p_\theta(Y_1^{k-1}) / p_\theta(Y_1^k)$ can be used to compute the marginal likelihood $p_\theta(Y) = p_\theta(Y_1^n)$ recursively.

\paragraph{Backward recursion.} 
The backward recursion is classical: we first observe that $\tau_{nq} = F_{nq}$, which is provided by the last step ($k = n$) of the forward recursion \eqref{eq:forward}. 
Then, for $n-1 \geq k \geq 1$, we obviously have $\tau_{kq} = \sum_{\ell=1}^Q \eta_{k+1, q, \ell}$ and the standard backward recursion for hidden Markov models writes 
\begin{align} \label{eq:backward}
\eta_{k+1, q, \ell} & = F_{kq} \; \pi_{q\ell} \; \frac{\tau_{k+1, \ell}}{G_{k+1, \ell}}, 
& \text{with} \quad
G_{k+1, \ell} & = \sum_{q=1}^Q  F_{kq} \pi_{q\ell}.
\end{align}

\subsection{Additional simulation results} \label{sec:addSimuls}

\paragraph{Accuracy of the parameter estimates.} ~

\begin{figure}[h]
  \begin{center}
    \includegraphics[width=.9\textwidth]{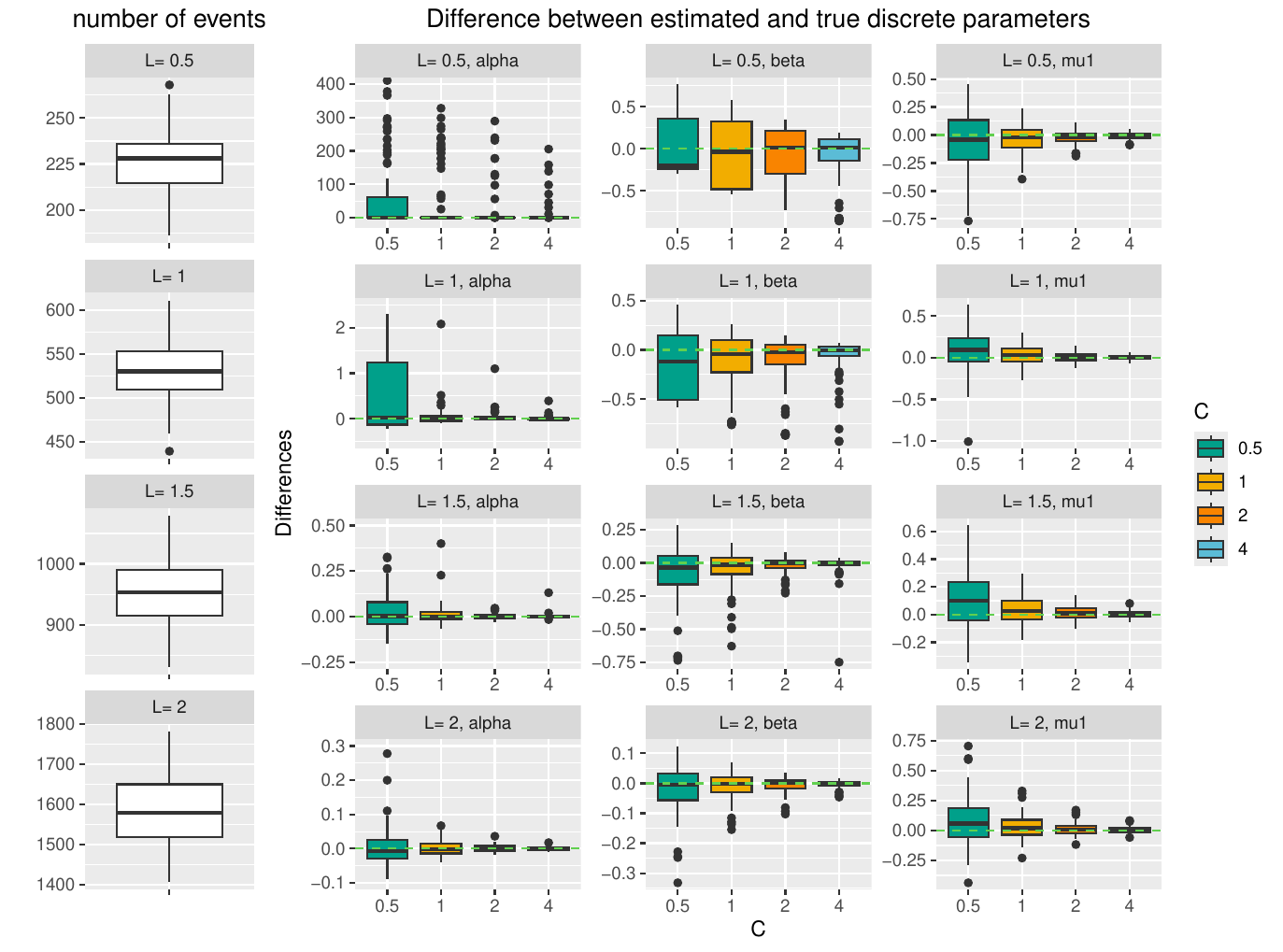}
    \caption{Accuracy of the estimates of the discrete parameters $(\alpha, \beta, \mu)$ for $Q = Q^* = 1$. 
    Same legend as Figure \ref{fig:discParmQ3}. 
    \label{fig:discParmQ1}}
  \end{center}
\end{figure}

\begin{figure}[h]
  \begin{center}
    \includegraphics[width=.9\textwidth]{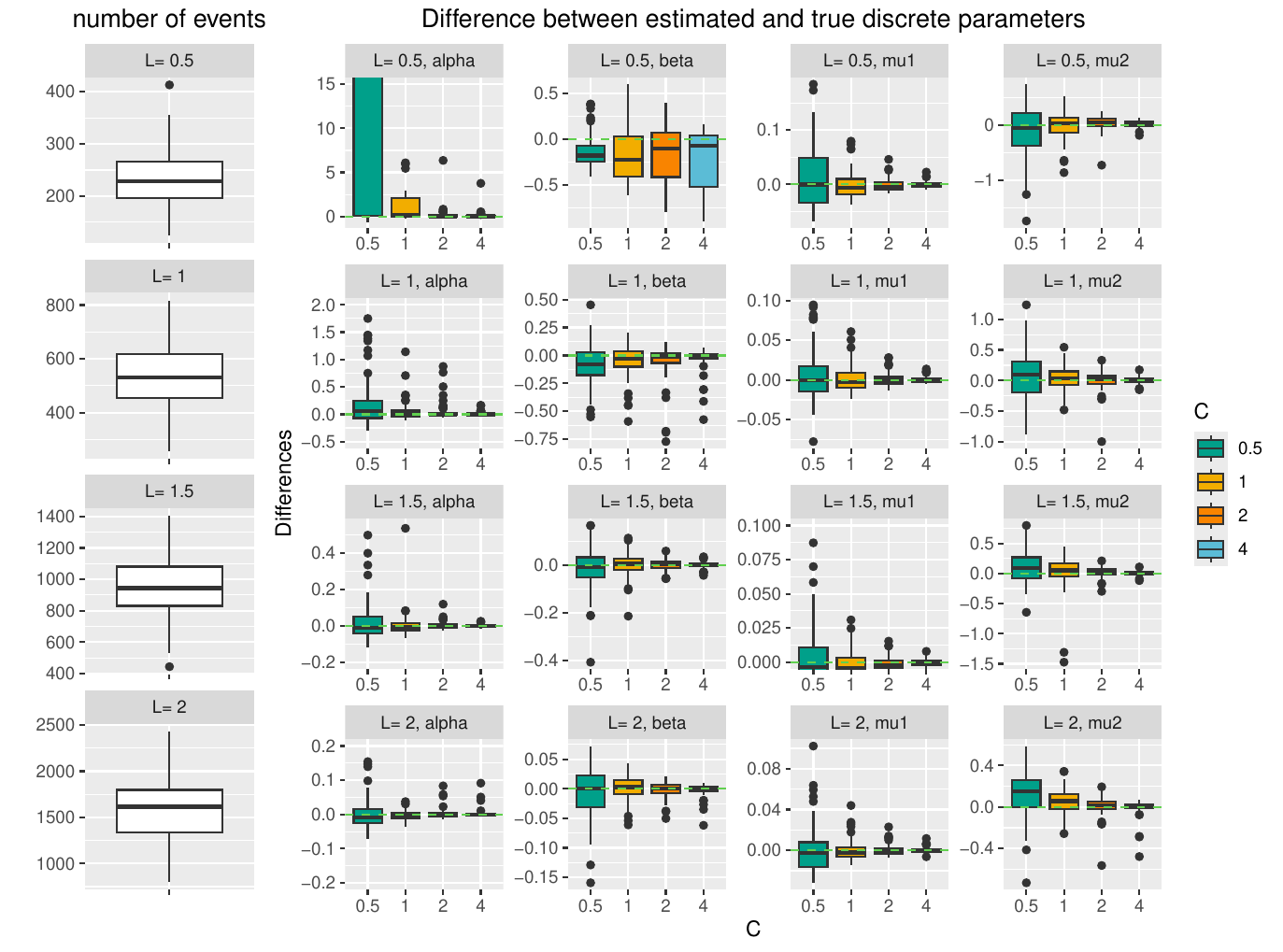}
    \caption{Accuracy of the estimates of the discrete parameters $(\alpha, \beta, \mu)$ for $Q = Q^* = 2$. 
    Same legend as Figure \ref{fig:discParmQ3}. 
    \label{fig:discParmQ2}}
  \end{center}
\end{figure}

\FloatBarrier

\paragraph{Model selection.} ~

\begin{figure}[h]
  \begin{center}
    \includegraphics[width=.9\textwidth]{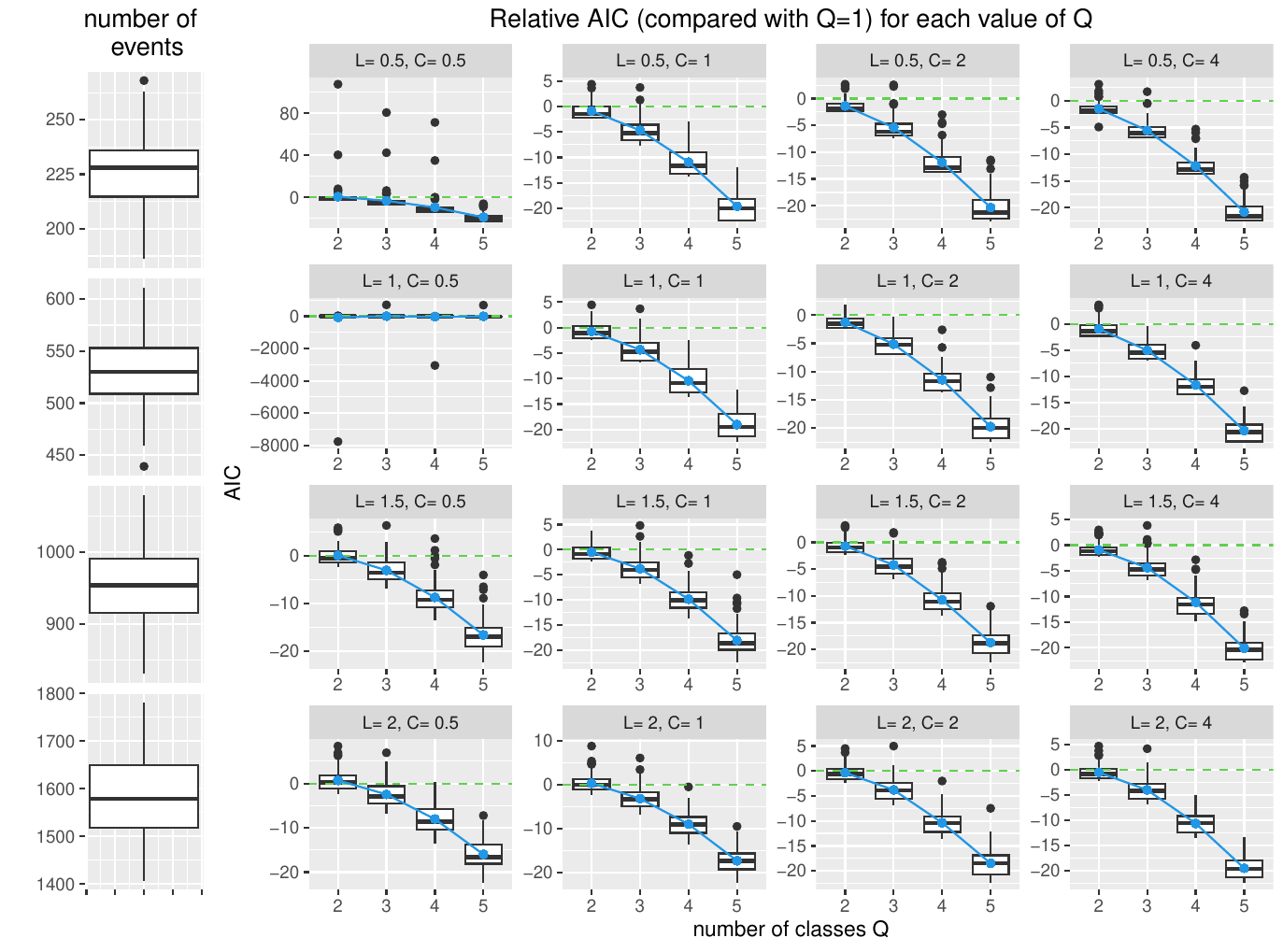}
    \caption{AIC criterion for the selection of $Q$ for $Q^* = 1$. 
    Same legend as Figure \ref{fig:aicQ3}. 
    \label{fig:aicQ1}}
  \end{center}
\end{figure}

\begin{figure}[h]
  \begin{center}
    \includegraphics[width=.9\textwidth]{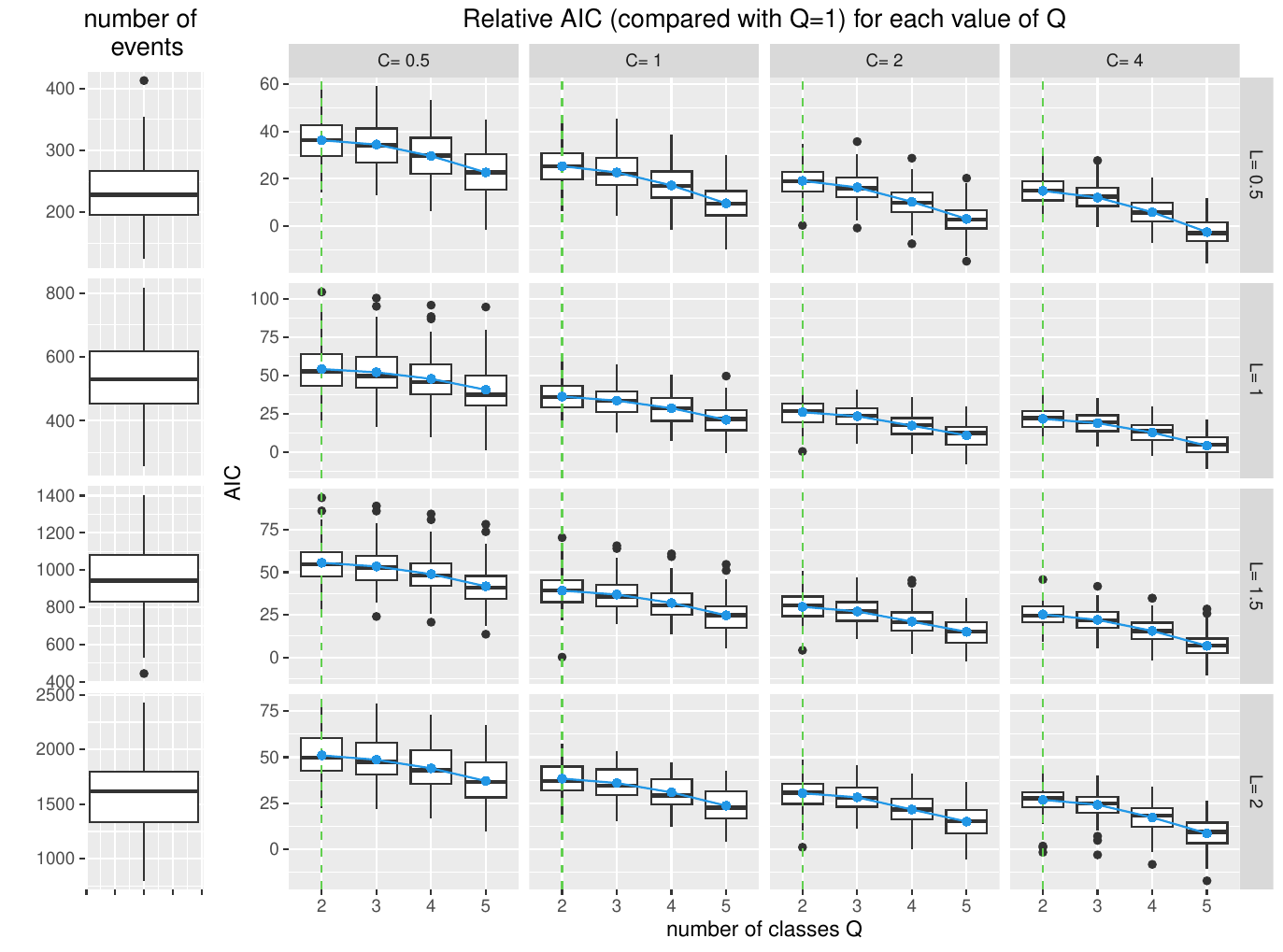}
    \caption{AIC criterion for the selection of $Q$ for $Q^* = 2$. 
    Same legend as Figure \ref{fig:aicQ3}. 
    \label{fig:aicQ2}}
  \end{center}
\end{figure}

\begin{figure}[h]
  \begin{center}
    \includegraphics[width=.9\textwidth]{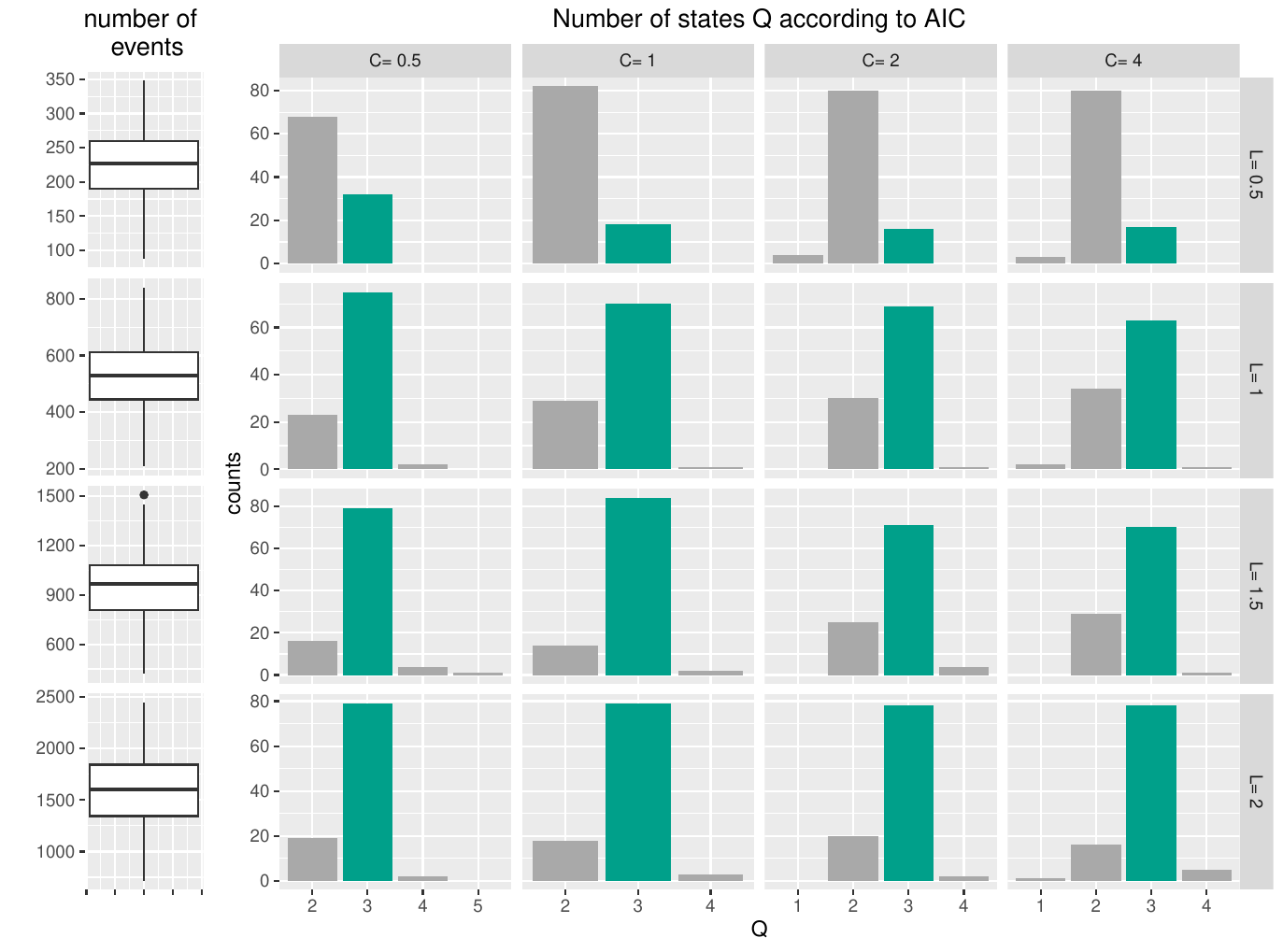}
    \caption{Estimated value of $Q$ according to AIC. The barplot is green when the estimated value corresponds to the true value $Q^* = 3$ and grey otherwise. Each barplot is obtained for a value of $L \in \{0.5, 1, 1.5, 2\}$ (row) and $C \in \{0.5,1,2,4\}$ (column).}
    \label{fig:Qaic_Q3}
  \end{center}
\end{figure}

\begin{figure}[h]
  \begin{center}
    \includegraphics[width=.9\textwidth]{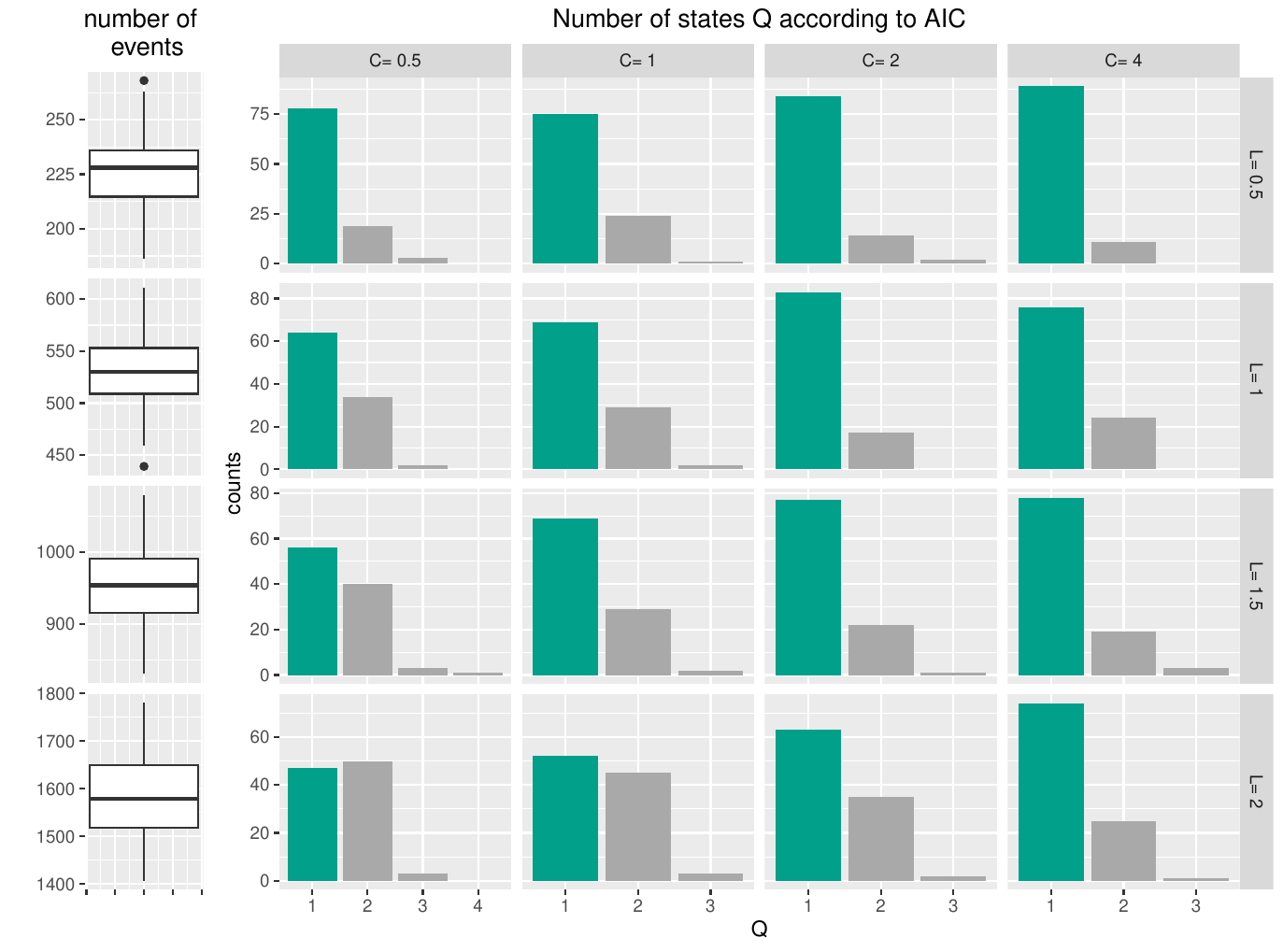}
    \caption{Estimated value of $Q$ according to AIC. The barplot is green when the estimated value corresponds to the true value $Q^* = 3$ and grey otherwise. Each barplot is obtained for a value of $L \in \{0.5, 1, 1.5, 2\}$ (row) and $C \in \{0.5,1,2,4\}$ (column).}
    \label{fig:Qaic_Q1}
  \end{center}
\end{figure}

\begin{figure}[h]
  \begin{center}
    \includegraphics[width=.9\textwidth]{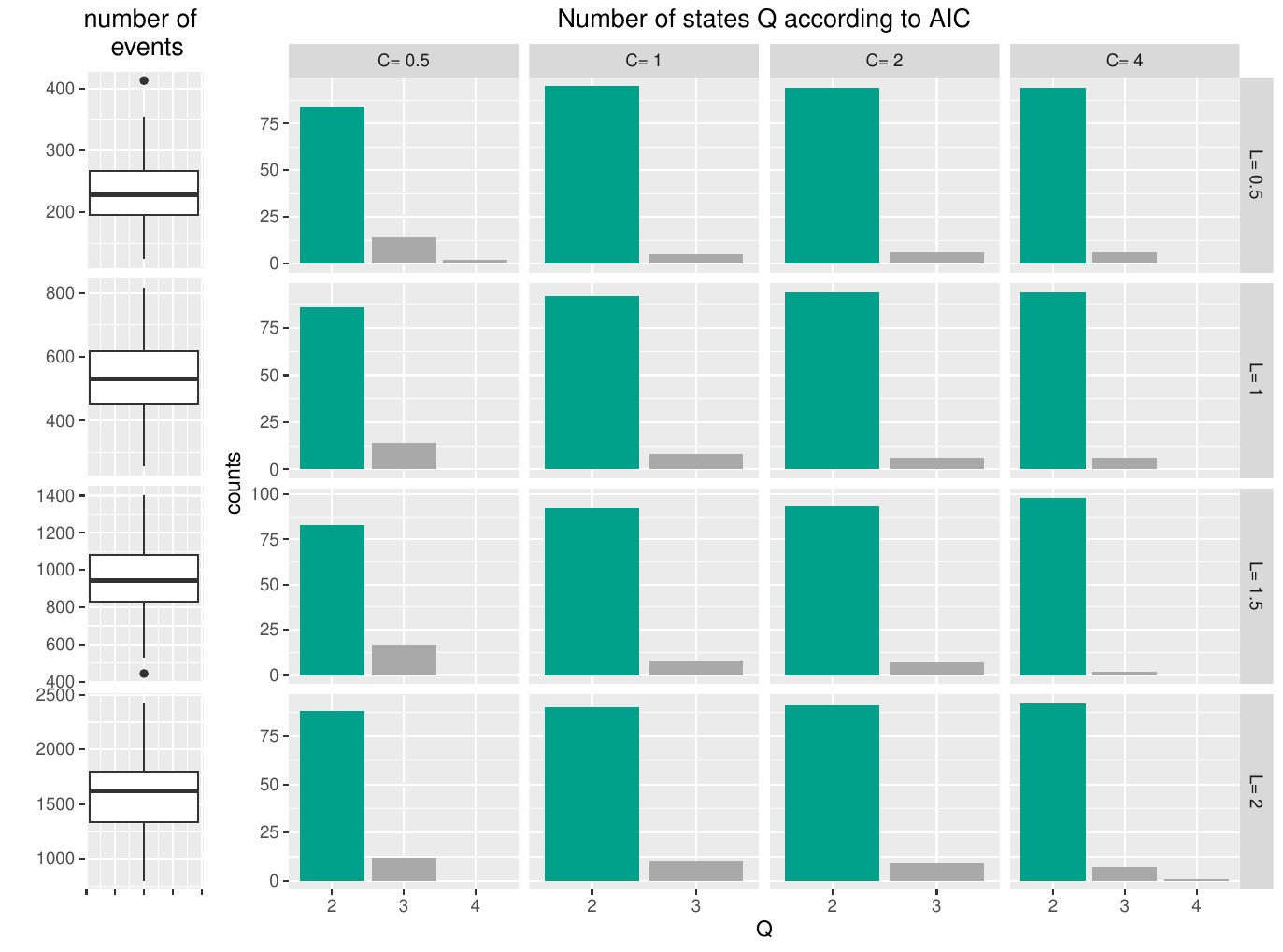}
    \caption{Estimated value of $Q$ according to AIC. The barplot is green when the estimated value corresponds to the true value $Q^* = 3$ and grey otherwise. Each barplot is obtained for a value of $L \in \{0.5, 1, 1.5, 2\}$ (row) and $C \in \{0.5,1,2,4\}$ (column).}
    \label{fig:Qaic_Q2}
  \end{center}
\end{figure}

\FloatBarrier

\paragraph{Classification and computational time.} ~

\begin{figure}[h]
  \begin{center}
    \includegraphics[width=\textwidth]{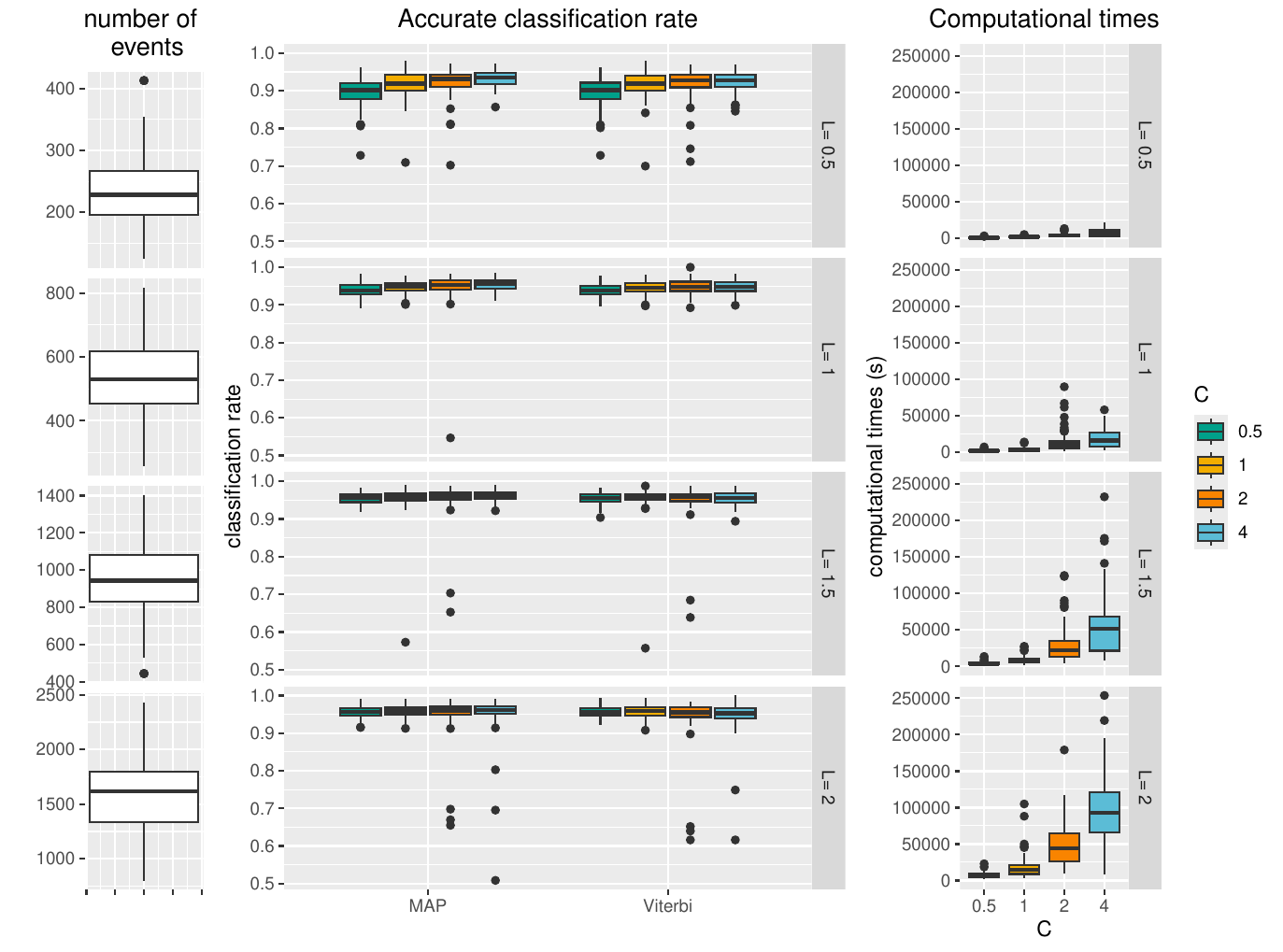}
    \caption{Proportion of well classified time bins (and computational time) for $Q^* = 2$ as a function of the discretization coefficient $C$.
    Same legend as Figure \ref{fig:diag-cpu-Q3}.
    \label{fig:diag-cpu-Q2}}
  \end{center}
\end{figure}

\FloatBarrier

\end{document}